\documentstyle%[referee,l-aa-fix]
{l-aa}
\input{psfig}

\makeatletter
\let\@internalcite\cite
\def\cite{\def\astroncite##1##2{##1 ##2}\@internalcite}
\def\citeyear{\def\astroncite##1##2{##2}\@internalcite}
\def\citename{\def\astroncite##1##2{##1}\@internalcite}
\def\nocite{\def\astroncite##1##2{}\@internalcite}
\def\citeasnoun{\def\astroncite##1##2{##1 (##2)}\@internalcite}
\def\possesivcite{\def\astroncite##1##2{##1's (##2)}\@internalcite}

\def\@normalsize{\@setsize\normalsize{12pt}\xpt\@xpt
\abovedisplayskip=3 mm plus6pt minus 4pt
\belowdisplayskip \abovedisplayskip
\abovedisplayshortskip=0mm plus6pt
\belowdisplayshortskip=2 mm plus4pt minus 4pt
\def\@listi{\topsep 4pt plus 2pt minus 2pt
\leftmargin \leftmargini
\parsep 0pt
\itemsep \parsep}}
\def\small{\@setsize\small{11pt}\ixpt\@ixpt
\abovedisplayskip 8.5pt plus 3pt minus 4pt
\belowdisplayskip \abovedisplayskip
\abovedisplayshortskip \z@ plus 2pt
\belowdisplayshortskip 4pt plus2pt minus 2pt
\def\@listi{\topsep 3pt plus 2pt minus 2pt
\leftmargin \leftmargini
\parsep 0pt
\itemsep \parsep}}

\def\scriptsize{\@setsize\scriptsize{8pt}\viipt\@viipt}
\def\tiny{\@setsize\tiny{6pt}\vpt\@vpt}
\def\large{\@setsize\large{17pt}\xivpt\@xivpt}
\def\Large{\@setsize\Large{17pt}\xivpt\@xivpt}
\def\LARGE{\@setsize\LARGE{20pt}\xviipt\@xviipt}
\def\huge{\@setsize\huge{25pt}\xviipt\@xviipt}
\def\Huge{\@setsize\Huge{30pt}\xviipt\@xviipt}
\@normalsize\rm

\def\maketitle{\par
 \begingroup
 \def\thefootnote{\fnsymbol{footnote}}
 \if@twocolumn
   \onecolumn \@maketitle
 \else
   \newpage \@maketitle
 \fi
 \global\@topnum\z@
 \thispagestyle{empty}\@thanks
 \def\protect{\noexpand\protect\noexpand}
 \let\inst=\@gobble
 \let\fnmsep=\@gobble
 \let\thanks=\@gobble
 \xdef\@title{\@title\unskip}
 \def\stripauthor##1\and##2\endauthor{%
    \xdef\@author{##1\unskip\unskip\if!##2!\else\ et al.\fi}}
 \expandafter\stripauthor\@author\and\endauthor
 \endgroup
 \setcounter{footnote}{0}
 \setbox0=\hbox{\small\rm\@author\unskip: \@title\unskip}
 \dimen@=\wd0\relax
 \advance\dimen@ by 2cm
 \ifdim\dimen@>\textwidth
    \typeout{^^JAandA Warning: The running head built automatically from
             \string\author\space and \string\title
             ^^Jexceeds the pagewidth, please supply a shorter form
             ^^Jusing \string\markboth\string{...\string}\string{...\string}
             after the \string\maketitle-command.}%
    \setbox0=\hbox{\small\rm Please give a shorter version with:
             {\tt\string\markboth\string{...\string}\string{...\string}}}%
 \fi
 \setbox\runheadbox=\hbox{\unhbox0}
 \def\@oddhead{\small\rm\hfil\copy\runheadbox\hfil\llap{\thepage}}
 \def\@evenhead{\small\rm\rlap{\thepage}\hfil\copy\runheadbox\hfil}
 \let\maketitle\relax
 \let\@maketitle\relax
 \gdef\@thanks{}\gdef\@author{}\gdef\@title{}\gdef\@subtitle{}%
 \let\thanks\relax}

\def\@maketitle{\newpage
 \rm\vbox to0pt{}\vskip-8mm
\if!\@headnote!\else {\LARGE \it
  \vskip 8.89mm
  \pretolerance=10000
  \rightskip=0pt plus 3cm
  \noindent\@headnote \par}\vskip -3.7mm\fi
 \vskip51.712mm
 {\LARGE \bf\boldmath
  \pretolerance=10000
  \rightskip=0pt plus 4cm
  \noindent\ignorespaces
  \centering \@title \par}\vskip .8cm
\if!\@subtitle!\else {\LARGE \bf\boldmath
  \vskip .05cm
  \pretolerance=10000
  \rightskip=0pt plus 3cm
  \centering\@subtitle \par}\vskip10pt\fi
 {\bf \vskip 1.5cm
\setbox0=\vbox{\setcounter{@auth}{1}\def\and{\stepcounter{@auth}}%
\def\thanks##1{}\@author\global\c@@inst=\c@@auth}%
\def\lastand{\ifnum\c@@inst=2\relax\unskip{} and \else
\unskip, and \fi}%
\setcounter{@auth}{1}%
\def\and{\stepcounter{@auth}\ifnum\c@@auth=\c@@inst\lastand\else
\unskip, \fi}%
 \centering\ignorespaces\large\@author\vskip1.54cm}
 {\rm
 \centering\institutename}
 \vfill
 \centering{\normalsize\it submitted to Astronomy \& Astrophysics}
 \par
 \vskip 2.0cm}

\def\abstract{
\begin{list}{}{\leftmargin=1.21cm
\rightmargin=1.21cm \labelwidth=0.0pt}
\item[\hskip\labelsep{\bf Abstract.}]}
\def\endabstract{\end{list}
\vskip3ptplus12pt\null\twocolumn}

\def\@sect#1#2#3#4#5#6[#7]#8{\ifnum #2>\c@secnumdepth
     \def\@svsec{}\else
     \refstepcounter{#1}\edef\@svsec{\csname the#1\endcsname\ }\fi
     \@tempskipa #5\relax
      \ifdim \@tempskipa>\z@
 \begingroup #6\relax
   \@hangfrom{\hskip #3\relax\@svsec}{\interlinepenalty \@M #8\par}
 \endgroup
       \csname #1mark\endcsname{#7}\addcontentsline
  {toc}{#1}{\ifnum #2>\c@secnumdepth \else
        \protect\numberline{\csname the#1\endcsname.}\fi
      #7}\else
 \def\@svsechd{#6\hskip #3\@svsec #8\csname #1mark\endcsname
        {#7}\addcontentsline
      {toc}{#1}{\ifnum #2>\c@secnumdepth \else
        \protect\numberline{\csname the#1\endcsname.}\fi
         #7}}\fi
     \@xsect{#5}}

\makeatother

\def\dot{\mathaccent"795F }

\newcommand{\bref}[1]{(\ref{#1})}               % \ref with parenthesis
\newcommand{\geff}{$ g_\ast$}

\newcommand{\tnu}{$\tau$  neutrino}
\newcommand{\tnus}{$\tau$  neutrinos}

\newcommand{\tot}{{\rm d}}
\renewcommand{\phi}{\varphi}
\renewcommand{\rho}{\varrho}
\newcommand{\lsim}{\hbox{\raise.5ex\hbox{$<$}
    \kern-1.em\lower.6ex\hbox{$\textstyle{\sim}$}}} % ungefaehr kleiner
\newcommand{\gsim}{\hbox{\raise.5ex\hbox{$>$}
    \kern-1.em\lower.6ex\hbox{$\textstyle{\sim}$}}} % ungefaehr groesser
\def\he#1{{\rm{^{#1}He}}}
\def\li#1{{\rm{^{#1}Li}}}

\begin{document}

\thesaurus{2(02.05.1;02.14.1;12.05.1)}
 
\title{Primordial nucleosynthesis with massive \tnus}

\author{J. B. Rehm \inst{1}
\and  G. G. Raffelt \inst{2}
 \and A. Weiss \inst{1}}

\offprints{J. Rehm (jan@mpa-garching.mpg.de)}

\institute{Max-Planck-Institut f\"ur Astrophysik,
Karl-Schwarzschild-Str.1, 85745 Garching, Germany
\and Max-Planck-Institut f\"ur Physik, F\"ohringer Ring 6, 80805
M\"unchen, Germany}

\date{Received; accepted}

\maketitle

\begin{abstract}
  A massive long-lived \tnu\ in the MeV regime modifies the primordial
  light-element abundances predicted by big-bang nucleosynthesis (BBN)
  calculations. This effect has been used to derive limits on
  $m_{\nu_\tau}$. Because recently the observational situation has
  become somewhat confusing and, in part, intrinsically inconsistent,
  we reconsider the BBN limits on $m_{\nu_\tau}$. To this end we use
  our newly developed BBN code to calculate the primordial abundances
  as a function of $m_{\nu_\tau}$ and of the cosmic baryon density
  $\eta$.  We derive concordance regions in the
  $\eta$-$m_{{\nu}_{\tau}}$-plane for several sets of alleged
  primordial abundances. In some cases a concordance region exists
  only for a nonvanishing $m_{{\nu}_{\tau}}$. At the present time BBN
  does not provide clear evidence for or against a \tnu\ mass.
\end{abstract}

%%%%%%%%%%%%%%%%%%%%%%%%%%%%%%%%%%%%%%%%%%%%%%%%%%%%%%%%%%%%%%%%%%%%%%%%%%%%%%%
%% Main Text  %%%%%%%%%%%%%%%%%%%%%%%%%%%%%%%%%%%%%%%%%%%%%%%%%%%%%%%%%%%%%%%%%
%%%%%%%%%%%%%%%%%%%%%%%%%%%%%%%%%%%%%%%%%%%%%%%%%%%%%%%%%%%%%%%%%%%%%%%%%%%%%%%
\section{Introduction}

The theory of big-bang nucleosynthesis (BBN) describes the creation of
the light elements hydrogen, helium, lithium, and beryllium in the
early universe.  The standard version of this theory is based on three
main assumptions. 1.~The evolution of the universe is described by the
homogeneous and isotropic Friedmann-Robertson-Walker model. 2.~The
standard model of particle physics is used.  3.~The lepton asymmetry
of the universe is of the same order as the baryon asymmetry so that
the chemical potential of the neutrinos in the early universe is small
relative to the temperature.
\begin{figure}
\vspace*{-.8cm}
\centerline{\psfig{figure=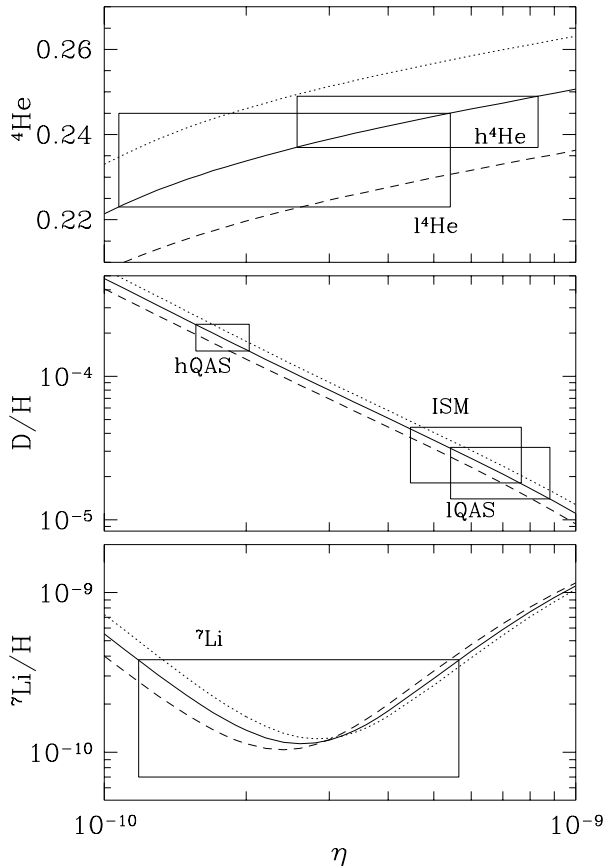,height=12cm}} %width=\columnwidth}}
%\picplace{9cm}
\caption{Predicted values for the primordial abundances of the light
elements assuming 2 (dashed), 3 (solid), and 4 (dotted) light neutrino
families. The boxes indicate observationally inferred
limits (Sect.~\protect\ref{obs}).
\label{f:SBBN}}
\end{figure}

Because the nuclear reaction rates and the neutron half-life are well
known, there remains only one free parameter, namely the present-day
cosmic number density $\eta$ of baryons relative to photons.  For a
given $\eta$ it is then possible to solve numerically a nuclear
network in the expanding universe which gives the primordial light
element abundances. In Fig.~\ref{f:SBBN} we show the results of such a
calculation for $10^{-10}<\eta<10^{-9}$, based on our newly developed
BBN-code which is documented in \citeasnoun{JR:96}.  The results agree
very well with those of other groups (e.g.~\cite{YTSSO:84};
\cite{SKM:93}).

\begin{table*}
\caption{Observationally inferred primordial light-element
abundances. ``Measured'' values and the errors are taken from the
given papers in which  references to the original observations
can be found. The errors are systematic except for those with the
superscript ``stat'' (for statistical).  For $\li7$ two sets of
systematic errors are given as explained in the text.}
\label{t:SBBN}
\begin{flushleft}
\begin{tabular}{lllll}
\hline
\hline
Abundance&Measurement&Adopted Range&Label&Reference\\
\hline
$Y_{\rm p}$
&$0.234\pm0.003^{\rm stat}\pm0.005$&0.223--0.245
&l$\he4$&\citeasnoun{OS:96}\\
&$0.243\pm0.003^{\rm stat}$&0.237--0.249
&h$\he4$&\citeasnoun{ITL:97}\\
${\rm D/H}\times 10^5$&$1.6\pm 0.2$
&1.8--4.4&ISM&\citeasnoun{LBG:93}, \citeasnoun{DST:96}\\
&$19\pm4$&15--23&hQAS&\citeasnoun{RH:96}\\
&$2.3\pm0.3^{\rm stat}\pm0.3$&1.4--3.2&lQAS&\citeasnoun{TFB:96}\\
$\li7/{\rm H}\times10^{10}$
&$1.6\pm0.07^{{\rm stat\,}+0.4+1.6}_{{\phantom {\rm stat\,}}-0.3-0.3}$
&0.9--3.7&$\li7$&\citeasnoun{FKOT:96}\\
\hline
\hline
\end{tabular}
\end{flushleft}
\end{table*}

In the past, BBN has become  a cornerstone of the hot big bang
scenario in that one could find a concordance interval for $\eta$
where all observed light-element abundances agreed with the calculated
values. Further, BBN has been used to constrain new particle-physics
models, notably the properties of neutrinos, and thus has become a
powerful tool in the field of particle astrophysics
(e.g.~\cite{MM:93}). In particular, 
if $m_{\nu_\tau}$, the mass of the \tnu, is larger than a few
$100\,\rm keV$ the cosmic expansion 
rate during the BBN epoch is modified enough to cause significant
changes in the predicted light-element abundances. Therefore,
one can derive constraints on a
putative mass for the $\tau$ neutrino or other neutral leptons that
might be present in the early universe (\cite{DKTW:78,MS:78,KS:82,DK:88}).

Since this earlier works, the $m_{\nu_\tau}$ question has been
revisited by many groups of authors   
(\cite{KTCS:91,DR:93,KKKSSW:94,DPV:96},
Hannestad \& Madsen 1996a,b, \cite{FKO:97}). The measured width of the $Z^0$ boson
shows that there are exactly three neutrinos which are lighter than
half the $Z^0$ mass so that the post-1990 works 
indeed
focus on the question of a $\tau$ neutrino mass rather than on the
number of light neutrino flavors or the mass of some arbitrary new
lepton. One motivation for the post-1990 works 
was to understand the impact of certain approximations for
the kinetic treatment of massive neutrino freeze-out. Another
motivation was to include the latest observationally inferred
primordial light-element abundances in the derivation of the
constraints. While the 
treatments by the various groups differ in detail, they all
agree that $m_{\nu_\tau}$ above a few $100\,\rm keV$ is forbidden. A
mass beyond a few $10\,\rm MeV$ would again be allowed by BBN, but is
experimentally ruled out (ALEPH--collaboration \cite{BCD:95,Pa:96}). 

Meanwhile, the observational situation has changed in that the first
direct measurements of deuterium in quasar absorption systems
have become available. Unfortunately, the results by
different authors 
and from different systems do not agree with each other, and do not
necessarily agree with the classic interstellar-medium determination
of the primordial abundance. In addition, a new determination of the
primordial $^4$He abundance yields a significantly higher value than
the previous ``standard'' result. Therefore, at the present time it is
not clear which set of observational data (if any of the currently
available ones) should be used to compare with the BBN calculations.
It is not even clear whether there  indeed exists  any concordance interval for
$\eta$, and if so, where exactly it lies. 
This constitutes the current debate about a possible ``BBN
crisis'' (see \cite{HSSTWBL:95,CST:95}).

Therefore, while 
the most recent papers on the $m_{\nu_\tau}$ question have
focused on various aspects of the kinetic treatment for these
particles during and after freeze-out, the main current
source of uncertainty for $m_{\nu_\tau}$ 
limits is actually the observational
situation. The main purpose of the present note is to
reconsider BBN limits on $m_{\nu_\tau}$ in the light of the current
debate about the correct primordial light-element abundances. In this
regard our study is motivated by a similar discussion of the allowed
number of massless neutrino species (\cite{CF:96}). 

These authors have calculated the primordial light-element abundances
as a function of the cosmic baryon-to-photon ratio $\eta$, and of the
equivalent number of massless neutrino species $N_{\rm eff}$ which
characterizes the energy density and thus the  expansion rate of the
universe at the time of BBN. They have then derived detailed 
concordance regions in the $\eta$-$N_{\rm eff}$-plane
where different sets of observational data agree with the calculated
abundances. However, these results cannot be trivially translated into
concordance regions in the $\eta$-$m_{\nu_\tau}$-plane because 
the contribution of a massive tau neutrino to the expansion rate
cannot be expressed in terms of an equivalent $N_{\rm
eff}$, as we will discuss below. 
Therefore, it is best to construct such concordance regions
directly in the \mbox{$\eta$-$m_{\nu_\tau}$-plane.} Before the onset of the 
current observational debate this sort of analysis 
was performed by \citeasnoun{KKKSSW:94}. 

A rather complete numerical treatment of the $\nu_\tau$ kinetics was
performed in the works of \citeasnoun{KKKSSW:94} and Hannestad \&
Madsen (1996a,b). Together with the recent studies by
 \citeasnoun{DPV:96} and \citeasnoun{FKO:97} they indicate that the final
concordance regions in the $\eta$-$m_{\nu_\tau}$-plane are rather
insensitive to fine points of the $\nu_\tau$ kinetics. Therefore, we
limit ourselves to the simplest approximation in which kinetic
equilibrium is assumed for $\nu_\tau$ throughout so that an
analytically integrated Boltzmann collision equation can be used.
With this approach we can reproduce previous results with sufficient
accuracy.

Of course, on a timescale much longer than the
BBN epoch (approximately 200 seconds), the \tnus\ have to decay into
relativistic particles to prevent the universe from becoming
overdominated by \tnus\ today (e.g. \cite{kt:90}). 
We will not discuss the impact of neutrino decays on
BBN. Extensive previous discussions include those by
\citeasnoun{KKKSSW:94} and \citeasnoun{DGT:94}.

We begin our study in Sect.~\ref{obs} with a short review of the
current status of the observations. In Sect.~\ref{s:massnu} we describe
the influence of a massive \tnu\ on the cosmic expansion rate, which
is calculated by solving a simplified Boltzmann equation. In
Sect.~\ref{s:abundmassnu} the influence of a massive \tnu\ on the
primordial abundances is studied.  We derive concordance regions in
the \mbox{$\eta$-$m_{\nu_\tau}$}-plane for different sets of observational
data.  Finally, we discuss and summarize our results in
Sect.~\ref{summ}. Some details about the neutrino reaction rates are
given in Appendix~\ref{s:rates}.

\section{Observations}
\label{obs}

In order to derive concordance intervals for the cosmic baryon
density, or in order to derive BBN limits on novel particle-physics
models, one must compare the calculated light-element abundances with
observations.  For most elements the difficulty is to derive the
primordial abundances from values which are measured today in our
vicinity. Even for $\he4$, where this appears to be relatively
straightforward and model-independent, some controversy has
emerged. Therefore, we begin with a discussion of the current
observational situation.

\subsubsection*{Helium-4}

A large number of measurements exist for the $\he4$ as well as for
the O and N abundances in low-metallicity HII regions.  Using O or N
as tracers, the $\he4$ abundance is extrapolated to zero metallicity
to obtain the primordial value. We quote the usual ``standard'' value
(``low $\he4$'' labeled l$\he4$ in Table~\ref{t:SBBN}) according to
\citeasnoun{OS:96}. Their results are very similar to those of
\citeasnoun{OS:95}; the values and errors differ at most by $2 \times
10^{-3}$. In contrast,
\citeasnoun{ITL:97} have found a significantly higher value
(h$\he4$ in Table~\ref{t:SBBN}) using a new set of emissivities and
collisional enhancement factors for the analysis of the observed
spectral lines and by using other selection criteria for the HII
regions. Their method and interpretation of the differences, in turn,
has been critiqued by 
\citeasnoun{OSS:96}, who  reconfirm the above ``low'' value.

We emphasize that we do not give more weight to
any of the two values; we merely quote what is currently discussed in
the literature. In view of the disagreement between the lower
deuterium value inferred from quasar observations (see below) 
it has been proposed
that the systematic uncertainties might be
substantially larger than commonly assumed (\cite{SG:95}). They argue that the
primordial mass fraction of $^4$He might be as high as 0.26.

\subsubsection*{Deuterium}

For deuterium, there is the canonical value  from
observations of deuterium 
in the solar system (\cite{Ge:93}) and the interstellar medium
(\cite{LBG:93,LW:96,PWLDA:97}) which provide a lower limit on the
primordial deuterium abundance. Based on the theory of
galacto-chemical evolution for deuterium, an upper limit---usually
about a factor of 2--3 higher (cf. \cite{Ed:94})---can be inferred.
The most recent analysis is 
that of \citeasnoun{DST:96} which is labeled ``ISM'' in
Table~\ref{t:SBBN}. 
In all canonical models of galacto-chemical evolution a
strong overproduction of $^3$He with respect to observations in the
solar system and the ISM  is found. As long as this problem remains
unsolved, limits based on  an analysis of the sum of deuterium and
$^3$He should not be used to constrain the primordial abundances. The
value quoted here relies on deuterium depletion only.
Non--canonical models for stellar evolution, including additional
mixing on the Red Giant Branch avoid a net production of  $^3$He in low mass
stars and might provide a solution to this problem (\cite{Ch:95,WWD:96}).

In addition, deuterium is the only case where one
believes to observe a primordial abundance directly. Deuterium lines
were found in extragalactic HII clouds which lie on the line of sight
to high redshift quasars (Quasar Absorption Systems or QAS).
Unfortunately, the values found in different systems by different
authors do not agree.  \citeasnoun{TFB:96} argue that the correct
value is one order of magnitude smaller (``low QAS value'' labeled lQAS in
Table~\ref{t:SBBN}) than the one originally found by
\citeasnoun{SCHR:94} and \citeasnoun{CRWCW:94}. Their high value
(hQAS in Table~\ref{t:SBBN}) was recently confirmed in
an analysis by \citeasnoun{RH:96}.  The high QAS deuterium abundance
would allow for an $\eta$ concordance interval at relatively low
values when combined with the lower $\he4$ abundance, but poses severe
problems for the theory of galacto-chemical evolution of deuterium and
$\he3$ (e.g.~\cite{St:94,GPFP:95}). 
Even the non-canonical models of \citeasnoun{Ch:95} and
\citeasnoun{WWD:96} cannot make the high deuterium value 
consistent with the present--day $^3$He observations.
It is argued by \citeasnoun{SCHR:94},
\citeasnoun{CRWCW:94}, and \citeasnoun{TFB:96} that the 
high QAS deuterium value is in fact not deuterium, but another
hydrogen cloud at a different position which mimics the deuterium
line, and that the systems used are not sensitive for abundances as
low as $10^{-5}$. This interpretation is rejected by
\citeasnoun{RH:96}.  Clearly, more high redshift quasar spectra are
needed to reveal the true primordial deuterium abundance.

\subsubsection*{Lithium-7}

The chemical evolution of lithium in stars is not well understood and
therefore the derivation of the primordial value is much more
uncertain. We quote here a recent value from \citeasnoun{FKOT:96},
based on the observations by \citeasnoun{MPB:95},  for
the so-called ``Spite plateau'', which is the minimal $\li7$ abundance
found in old Pop II halo stars of our galaxy. Since it is nearly
constant in stars with a surface temperature of more than 5500~K and
metallicities lower than about 5~\% solar, it is usually argued that
this is the primordial value.  Because the abundance depends on the
modeling of the stellar atmosphere, the error is dominated by
systematic effects (first set of systematic errors in Table~\ref{t:SBBN}).
Furthermore, it has been suggested that $\li7$ might have been
depleted significantly (\cite{DDK:90,PDD:92}), but the detection of
the more fragile isotope $\li6$ in two of these stars may be taken as
an indication against strong depletion (\cite{SFOSW:93}). The possible
depletion effect is represented by the second set of systematic errors
in the $\li7$ abundance.
\smallskip

\noindent 
The current observational evidence is summarized in Table~\ref{t:SBBN}.
For each observation we also give a range for the relevant abundance
which we have adopted to compare with the calculations below. The
adopted range has been derived by adding the stated systematic error
linearly with the $2\,\sigma$ statistical one. For
lithium, both systematic errors have been linearly added. These
procedures are somewhat arbitrary, but represent the common practice
in this field.

\begin{figure}
\centerline{\psfig{figure=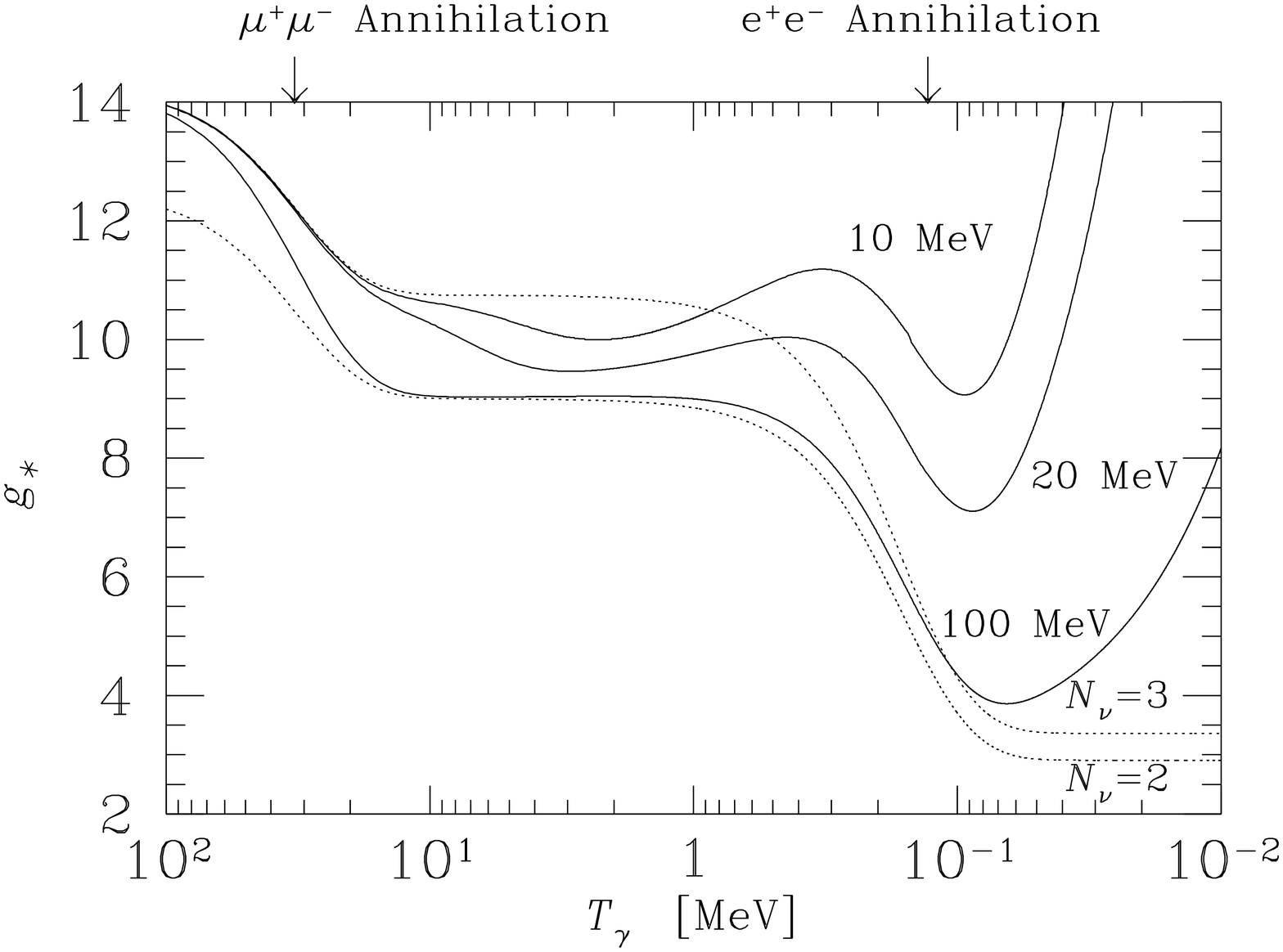,height=6.5cm}}
%\picplace{6.5cm}
\caption{Evolution of the parameter $g_*(t)$ for a Majorana \tnu\ with
$m_{\nu_\tau}=10$, 20, and $100\,\rm MeV$ (solid lines).  For
comparison, the evolution of \geff\ for 2 and 3 massless neutrino
families is also shown (dotted lines).
\label{f:gstmassnu}}
\end{figure}

\section{Massive long-lived \tnus\ in the early universe }
\label{s:massnu}

As a further ingredient for our study of possible BBN limits on
$m_{\nu_\tau}$ we need to specify the contribution of a massive
neutrino to the expansion rate of the early universe.  In the
radiation dominated phase the expansion rate is governed by the
Friedmann equation without a curvature term\footnote{We always use the
system of natural units where $\hbar=c=k_{\rm B} = 1$.}
\begin{equation}
H^2=\frac{8\pi}{3}\,\frac{\rho}{m_{\rm Pl}^2}.
\end{equation}
Here, $H=\dot{R}/R$ is the expansion rate, $m_{\rm Pl}$ the Planck
mass, and $\rho$ the energy density in radiation. It is commonly
expressed in the form
\begin{equation}\label{e:rho}
\rho= \frac{\pi^2}{30}g_* T^4
\end{equation}
where $T$ is the cosmic temperature and $g_*$ the effective number of
thermally excited relativistic degrees of freedom. This parameter is a
function of temperature and thus of cosmic time.  Just before the BBN
epoch one finds
\begin{equation}\label{e:gst}
g_*=g_\gamma+\textstyle{\frac{7}{8}}\left(g_{e} +  2 N_\nu\right)=10.75
\end{equation}
with $g_\gamma=2$ for the photons, $g_{e}=4$ for the electrons and
positrons, and $N_\nu=3$ for the number of neutrino families which are
assumed to be massless.  

A particle species no longer contributes when $T$ falls below its mass
because then its number density is suppressed and therefore the
contribution to the cosmic energy density in the early radiation
dominated phase of the universe is negligible.  The evolution of $g_*$
in the temperature range $10^2$ to $10^{-2}$ MeV is shown in
Fig.~\ref{f:gstmassnu} (dashed lines) for a scenario with 2 and 3
light neutrino families. At a temperature of about 1 MeV the electrons
get nonrelativistic and annihilate to photons. Therefore they drop out
of Eq.~\bref{e:gst}, $g_*$ decreases, and their entropy is transferred
to the photons, which dominate the energy density from now~on.

\begin{figure}
\centerline{\psfig{figure=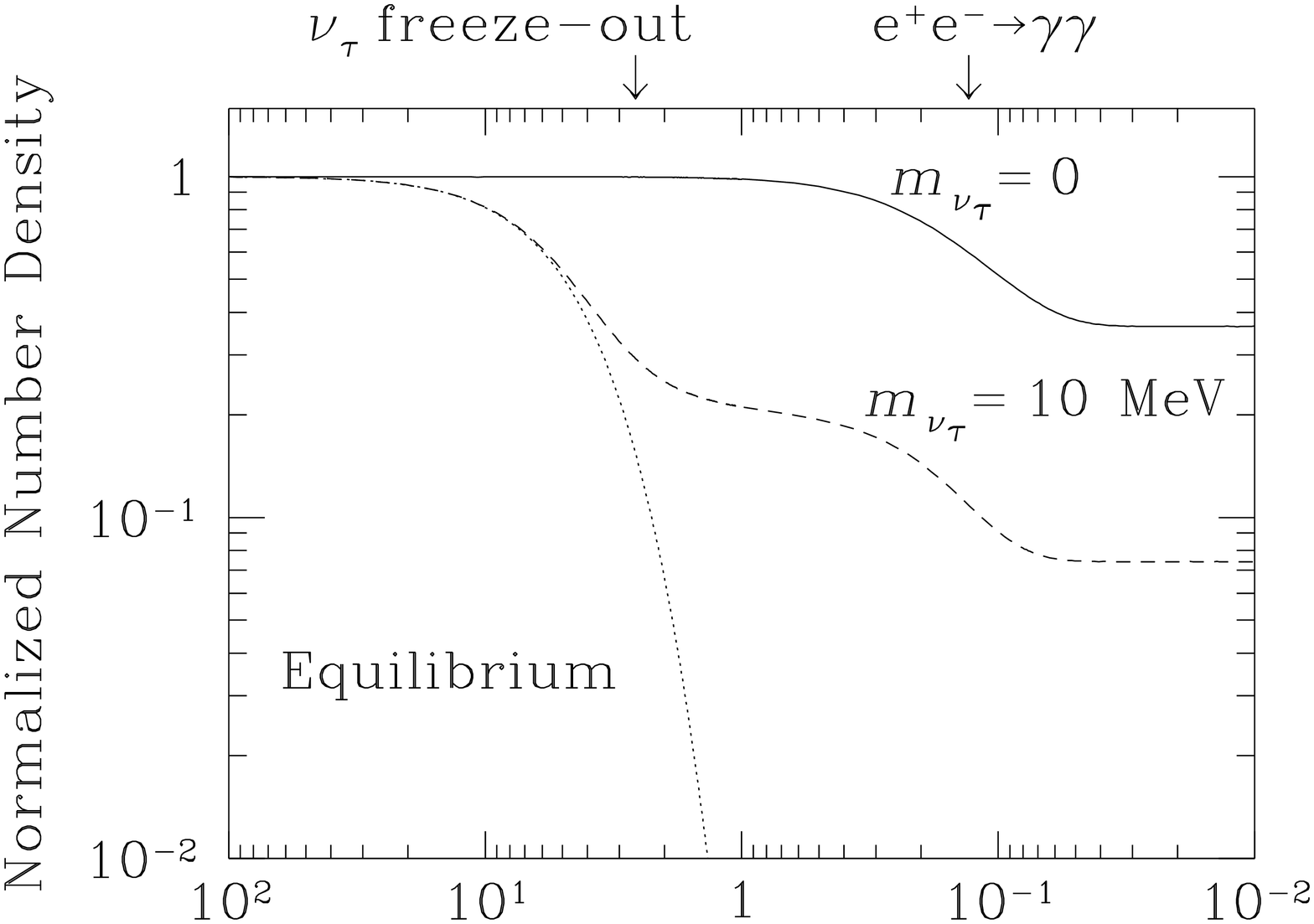,height=6.5cm}} %width=\columnwidth}}
\vskip-.5cm
\centerline{\psfig{figure=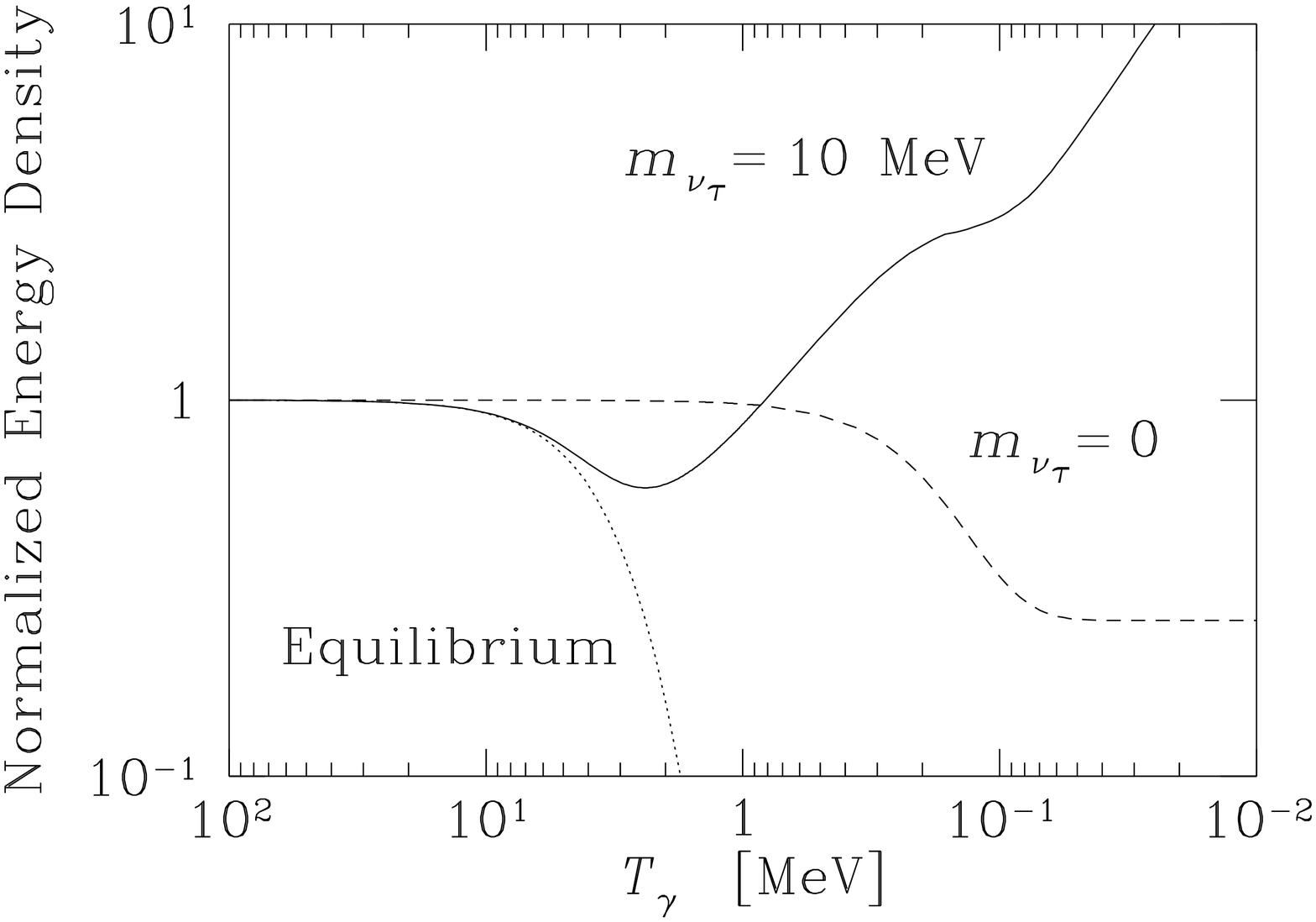,height=6.5cm}} %width=\columnwidth}}
%\picplace{8.25cm}
\caption{Number density (upper panel) and energy density
(lower panel) 
as a function of the cosmic temperature for a $10\,\rm MeV$ Majorana
\tnu. The contribution of a massless neutrino is shown for
comparison. The densities are relative to those of photons and
normalized to 1 at high temperatures. At
$e^+e^-$ annihilation the photon density increases and the relative
neutrino densities take a downward step as indicated. The curves
marked ``Equilibrium'' represent the $\nu_\tau$ densities if the
$\bar{\nu}_\tau\nu_\tau$ annihilations would not freeze
out.\label{massnu}}
\end{figure}

Massive neutrinos contribute differently to the cosmic energy density
than massless ones. Therefore, the expansion rate as a function of
temperature and thus the time-temperature relation is modified,
leading to changed light-element abundances in a BBN calculation.  We
examine a scenario where the \tnu\ is massive whereas the other two
neutrinos are taken to be massless. Thermodynamic equilibrium is
maintained as long as reactions such as
\begin{equation}
\nu_{\tau}+\bar{\nu}_{\tau}\longleftrightarrow a+\bar{a}
\end{equation}
are fast compared to the expansion of the universe. Here, $a$ stands
for any fermion which is kinematically allowed.  As the universe
expands, this reaction slows down and eventually the number density of
\tnus\ freezes out and stays constant thereafter
(Fig.~\ref{massnu}).

The process of freeze-out is described by the Boltzmann collision
equation. This integro-differential equation is very difficult to
solve in general, but it can be simplified by making the following
approximations: 1.~The annihilation products are in thermodynamic
equilibrium due to effective interactions with the cosmic
plasma. 2.~Classical instead of quantum statistics is used, i.e.\ a
Maxwell-Boltzmann distribution instead of a Fermi-Dirac or a
Bose-Einstein one.  3.~Even after annihilations the \tnus\ are kept in
kinetic equilibrium by effective elastic scatterings with electrons
and massless neutrinos which are far more abundant.  Other authors
have dropped some or all of these assumptions, but the results do not
change drastically, as we will discuss in Sect.~\ref{s:abundmassnu}.
Because of the assumed kinetic equilibrium, the distribution function
of the massive \tnus\ can be expressed in terms of the
so-called pseudo chemical potential~$z_i$,
\begin{equation}\label{distr}
f_{\nu_\tau}=\exp\left(z_i - \frac{E_i}{T}\right).
\end{equation}
Furthermore, the Boltzmann equation is integrated over the phase space
of one of the incoming \tnus. Because we assumed classical statistics,
this can be done in a closed form and one arrives at one single
ordinary differential equation for the $\nu_\tau$ number density,
\begin{equation}\label{BE}
        \frac{\tot n_{\nu_\tau}}{\tot t}+
3Hn_{\nu_\tau}=\left\langle\sigma v\right\rangle\;
\left[n_{\nu_\tau}(t)^2-n_{\nu_\tau}^{\rm eq}(t)^2
\right].
\end{equation}
Here, $H$ is the expansion rate, $n_{\nu_\tau}$ the actual and
$n_{\nu_\tau}^{\rm eq}$ the equilibrium number density of \tnus.  The
thermally averaged reaction rate $\langle\sigma v\rangle$ is given in
Appendix~\ref{s:rates}.
\begin{figure*}
\centerline{\psfig{figure=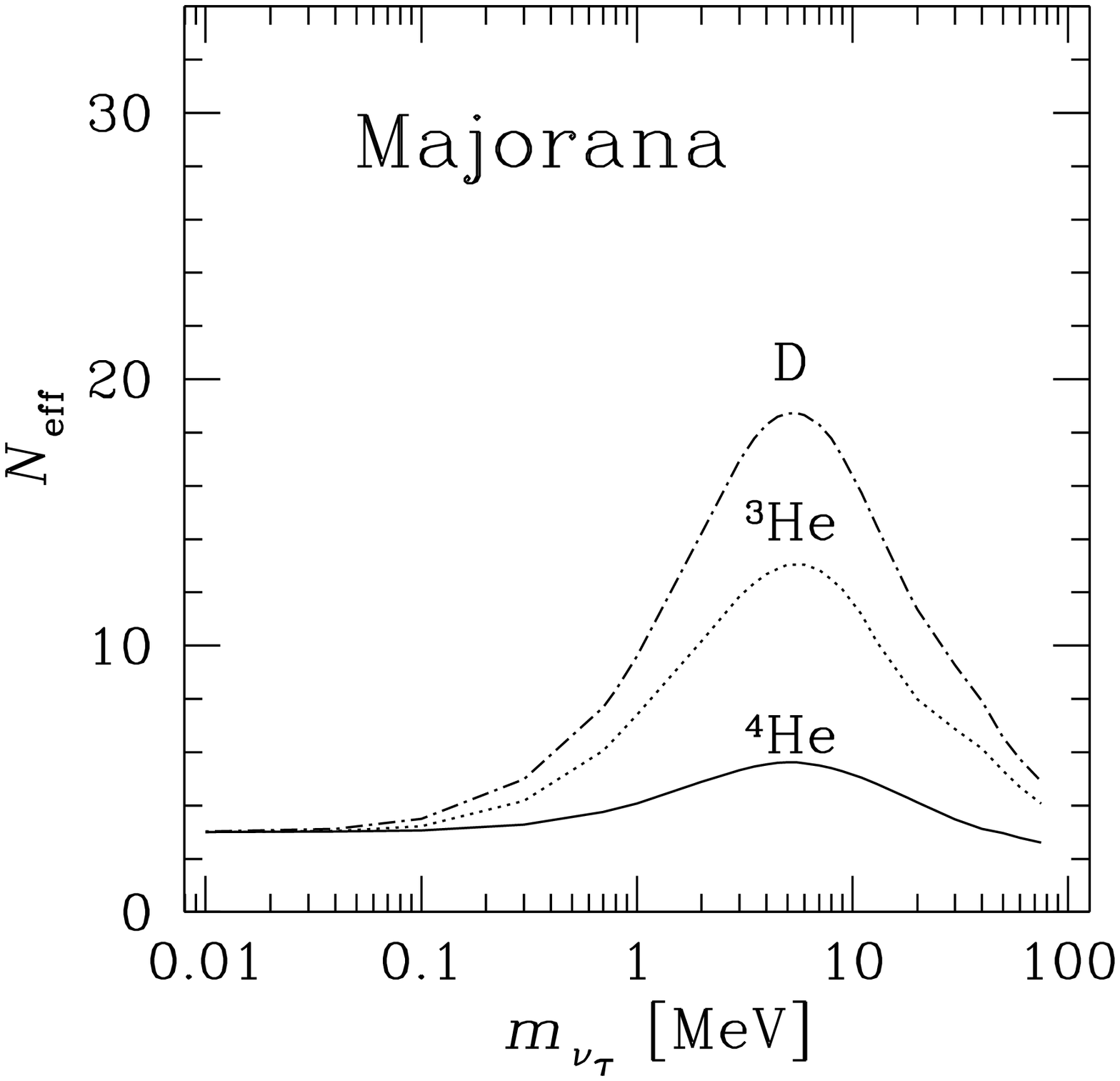,width=8.8cm}
\psfig{figure=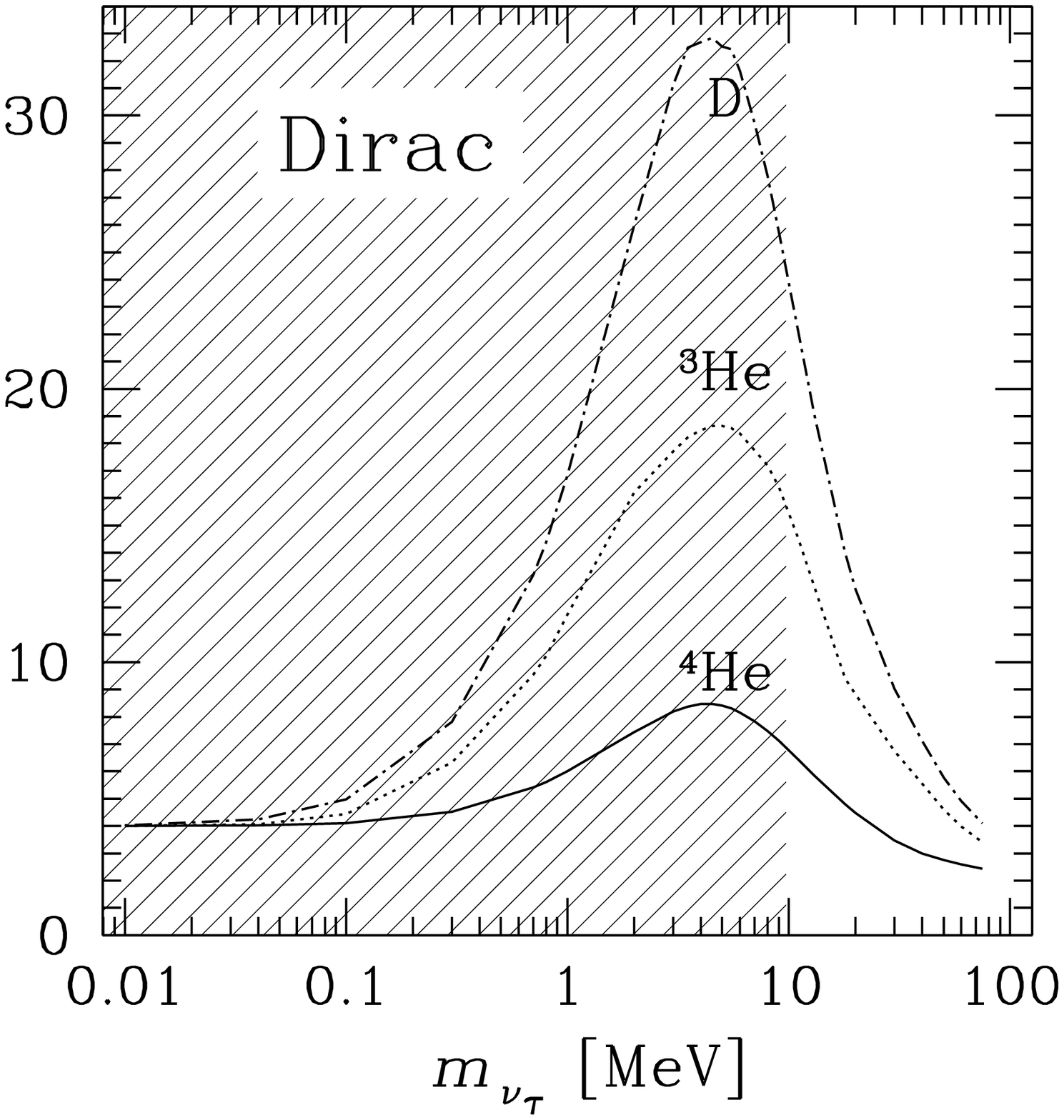,width=8.8cm}}
%\picplace{6cm}
\caption{Number of effective light neutrino families $N_{\rm eff}$
equivalent to the contribution of a massive neutrino. The value of
$N_{\rm eff}$ depends on the light element it is based upon. For Dirac
neutrinos, our ``helicity approximation'' causes $N_{\rm eff}$ to be
overestimated for $m_{\nu_\tau}\,\lsim\,10\, \rm MeV$; this region has
been hatched as a warning.
\label{f:nueff}}
\end{figure*}

For a Dirac neutrino we are confronted with another difficulty because
it has four degrees of freedom. In the massless case, only two of them
interact with other particles, namely the helicity-minus neutrino and
the helicity-plus antineutrino. The other two degrees of freedom are
sterile and have no impact at all.  If neutrinos have a mass, their
helicity eigenstates are no longer eigenstates of chirality and thus
are no longer completely sterile.  The strength of the weak
interaction for ``wrong-helicity'' neutrinos is suppressed by a factor
of about $(m_{\nu_\tau}/2E_{\nu_\tau})^2$ relative to the usual rates,
provided that $m_\nu\ll E_\nu$. 
In this case the quasi-sterile states remain essentially unexcited,
leaving us with two degrees of freedom which are relevant for the
cosmic expansion (for a recent discussion of cosmic abundances of
right-handed neutrinos with masses up to 200 keV
see \cite{EKMU:96}). 
Conversely, for nonrelativistic
neutrinos all four degrees of freedom interact with the full
strength. Therefore, if the neutrino mass exceeds about $1\,\rm MeV$
so that they are nonrelativistic at the time of freeze-out, all four
degrees of freedom would be fully excited and would thus contribute to
the expansion rate of the universe

A correct treatment of the intermediate-mass regime requires a
detailed kinetic treatment of neutrino helicities
(e.g.~\cite{FKO:97}).
 Because this is a very significant complication we have
limited ourselves to an approximation where it is assumed that all
available helicity states are always fully populated, an assumption
that we shall refer to as the ``helicity approximation''.  It implies
that the number and energy density of light ($m_\nu\lsim1\,\rm MeV$)
massive Dirac neutrinos is overestimated by a factor of 2. 
In the mass range from 1 -- 10 MeV the number and energy density is 
still overestimated, but now by a factor less than 2;
for $m_\nu\gsim10\,\rm MeV$ our approximation yields the correct result.

Because we have assumed kinetic equilibrium, the energy density of the
\tnus\ can be calculated from their number density.  As an example we
show in Fig.~\ref{massnu} the evolution of the number and energy
density, respectively, for a $10\,\rm MeV$ Majorana \tnu. Even though
the neutrino mass causes the universe to become quickly matter
dominated, we continue to parameterize the impact on the expansion rate
by $g_*$ which thus no longer represents the effective number of
relativistic degrees of freedom. It is merely a time-dependent
proportionality factor between $T^4$ and the energy density according
to the definition in Eq.~\bref{e:rho}.

\section{BBN with massive \tnus\ }
\label{s:abundmassnu}

\subsection{The number of effective light neutrino species}

It is common practice in the literature to parameterize the influence
on BBN of a new particle by the equivalent number $N_{\rm eff}$ of
light neutrino species.  This is a useful parameterization for a
particle which is relativistic during the entire period of BBN, i.e.\
down to $T\,\approx\, 40$ keV or until about 200 seconds after the big
bang. Such a particle contributes a fixed amount to the parameter
$g_*$ during the relevant epoch. On the other hand, a particle which
is or becomes nonrelativistic during the BBN epoch contributes to the
expansion rate in the way discussed in the previous section and displayed in
Fig~\ref{f:gstmassnu}, i.e.~its contribution to $g_*$ varies with
time. Evidently, the effect of a massive \tnu\
cannot be mapped on a constant shift of $g_*$ or on a fixed
modification of $N_{\rm eff}$. 

One can still define a useful $N_{\rm eff}$ parameter indirectly by
comparing the modification of the calculated abundance of a certain
light element, for example $^4$He, caused by a massive neutrino with
that caused by a fixed modification of $g_*$.  The
synthesis of the light elements peaks at different times for the
different elements. 
Therefore, defining $N_{\rm eff}$ in this way depends on the
chosen isotope. In Fig. \ref{f:nueff} we show $N_{\rm
eff}$ based on the equivalent impact on the abundances of deuterium
and the two helium isotopes.  (Note that in the Dirac case the
low-mass range does not converge to the correct value of 3 light
families, but to 4. This represents the abovementioned helicity
approximation.)

In the previous literature $N_{\rm eff}$ has
always been defined by the equivalent impact on $^4$He. However, if
one wishes to perform a more direct comparison with different sets of
observational data, there is no reason to introduce 
$N_{\rm eff}$. We will present all of our results directly
as a function of $m_{\nu_\tau}$ in a form analogous to that used by 
\citeasnoun{KKKSSW:94}. 

\begin{figure*}
\centerline{\psfig{figure=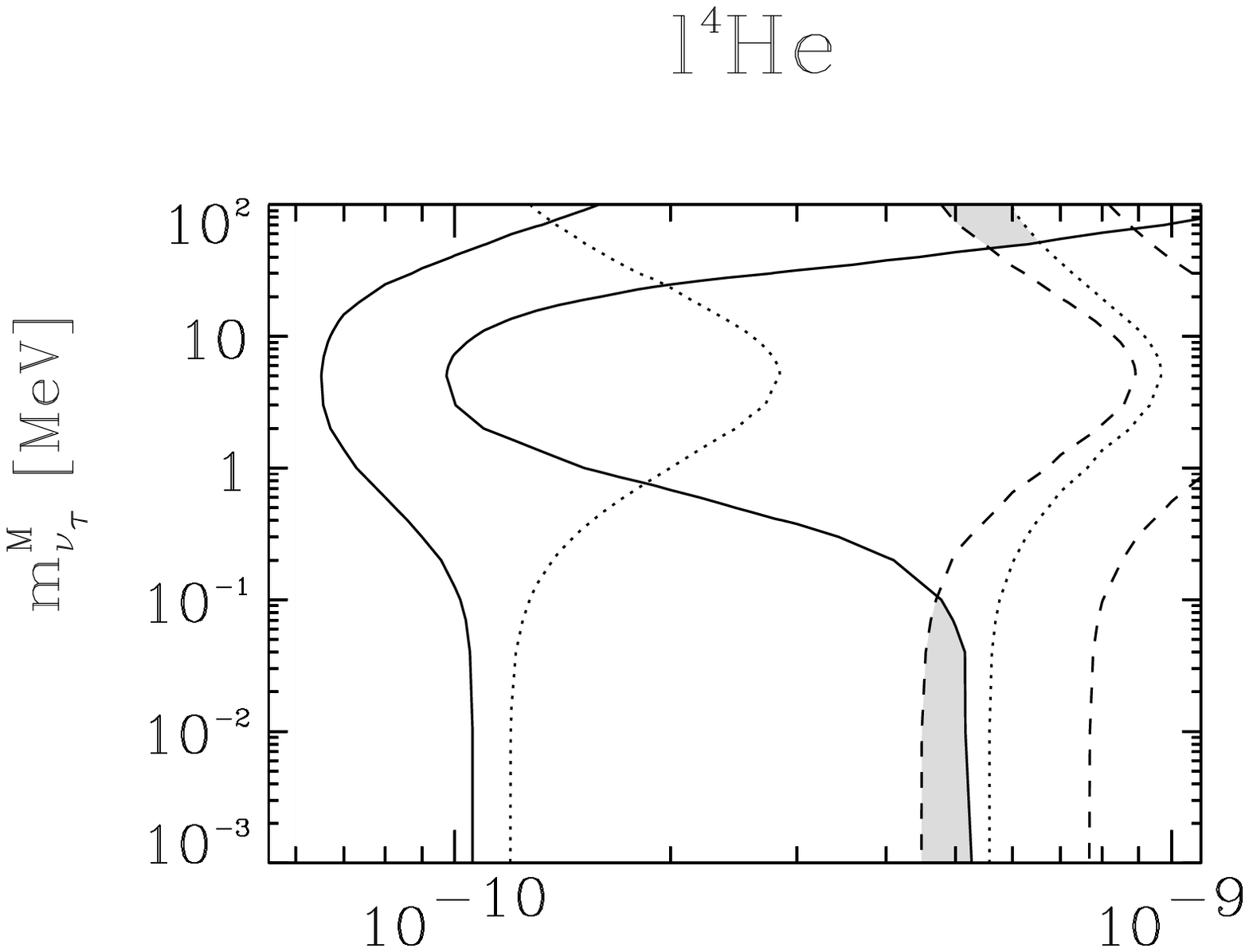,width=8.8cm}
\psfig{figure=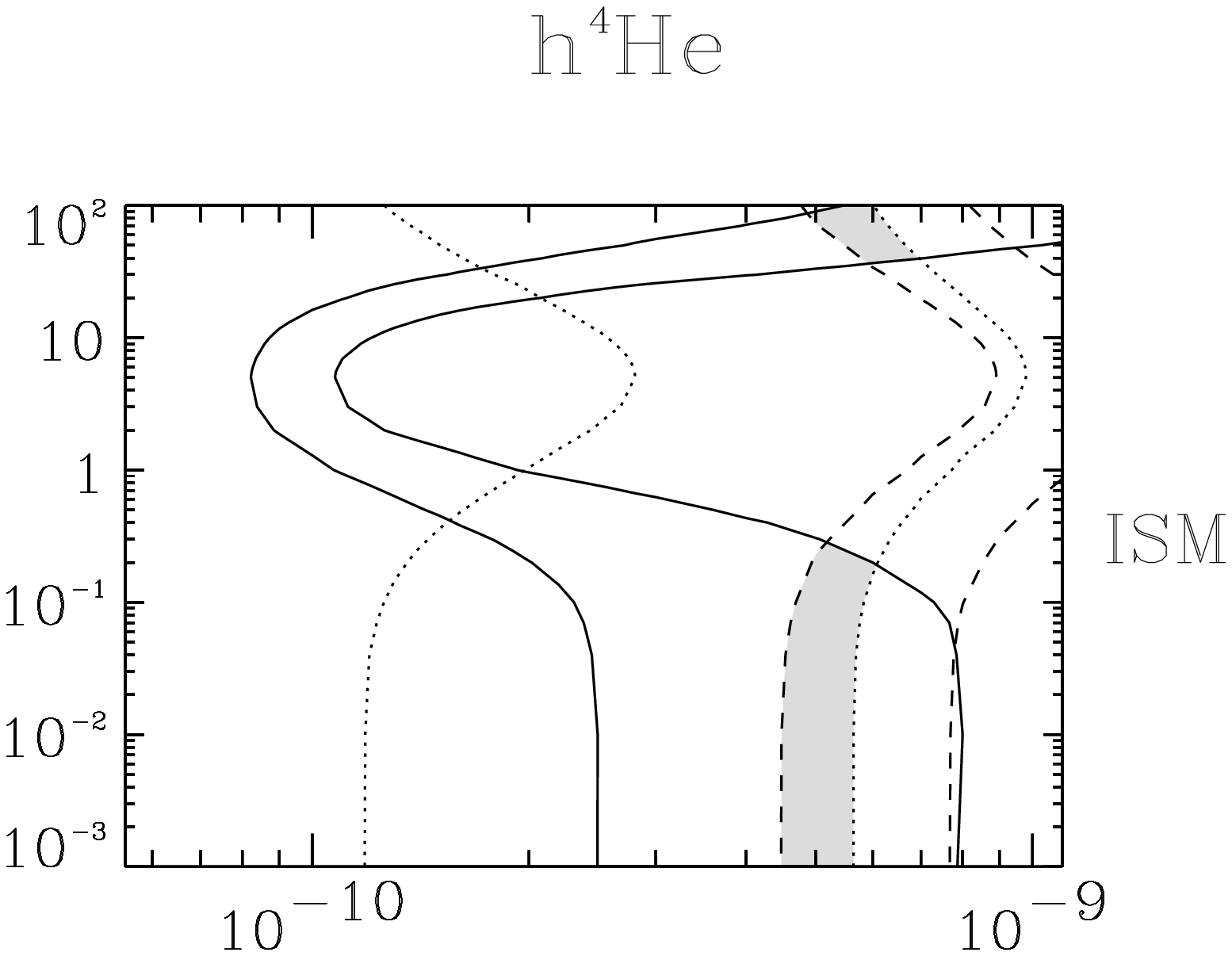,width=8.8cm}}
\vskip-1cm
\centerline{\psfig{figure=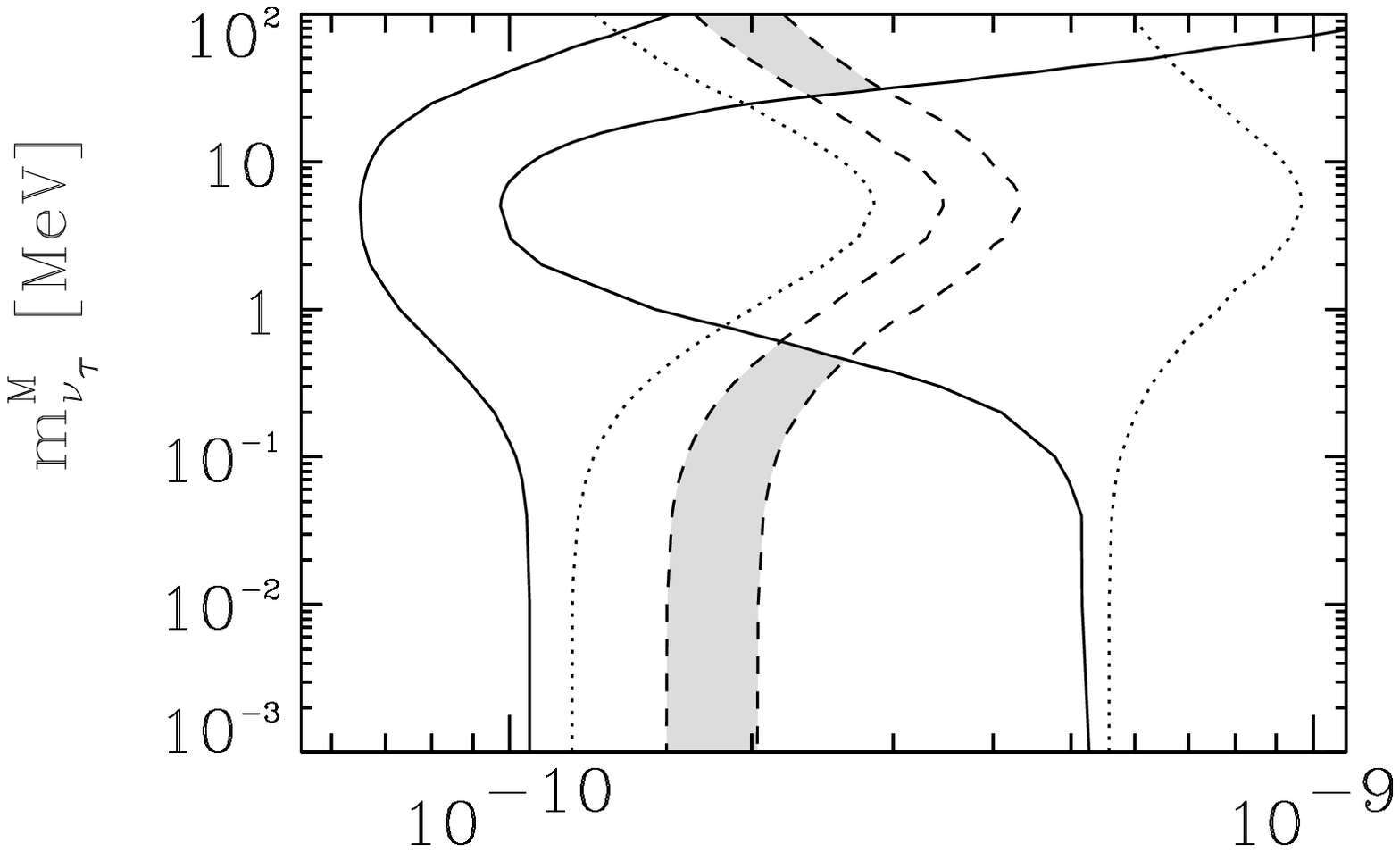,width=8.8cm}
\psfig{figure=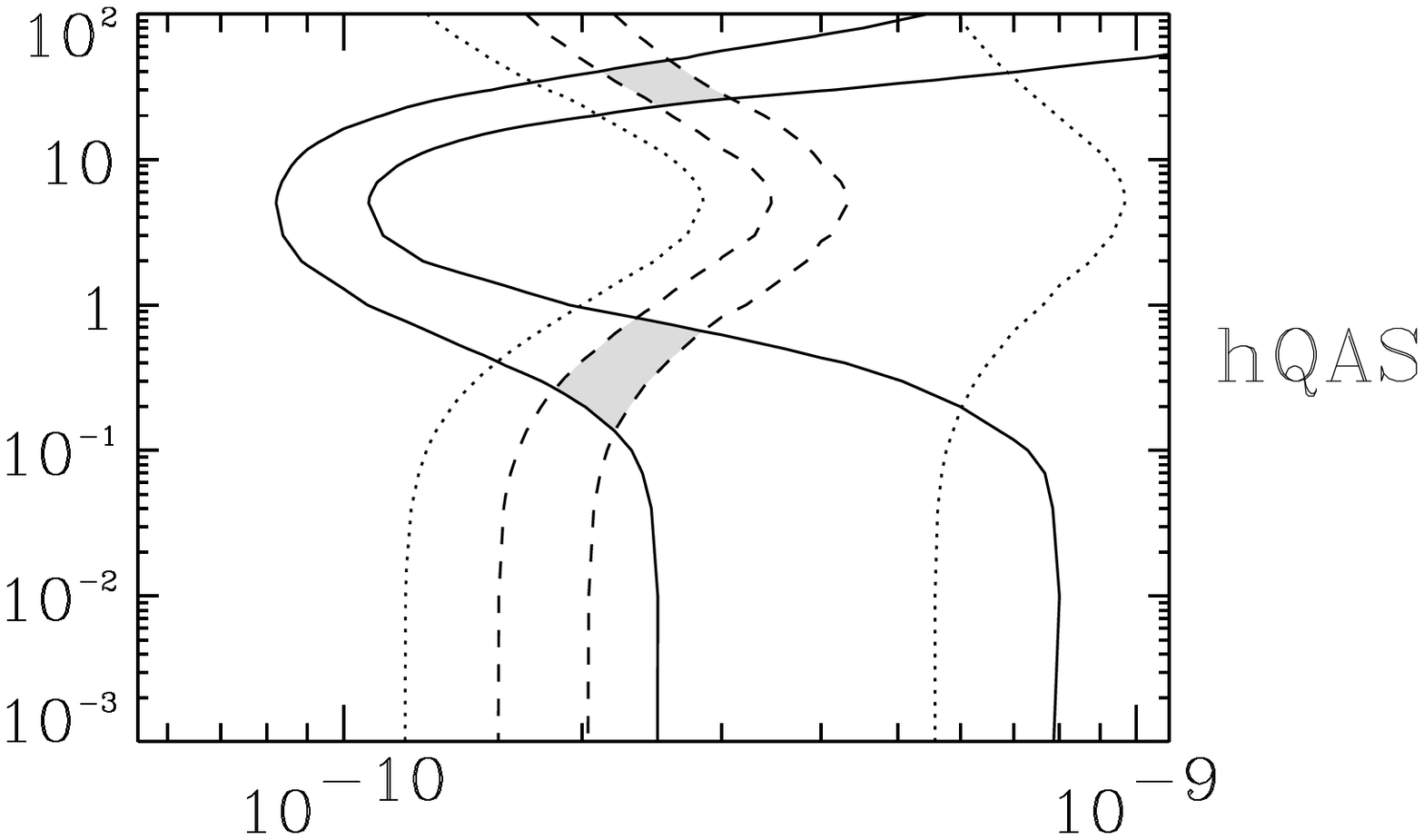,width=8.8cm}}
\vskip-1cm
\centerline{\psfig{figure=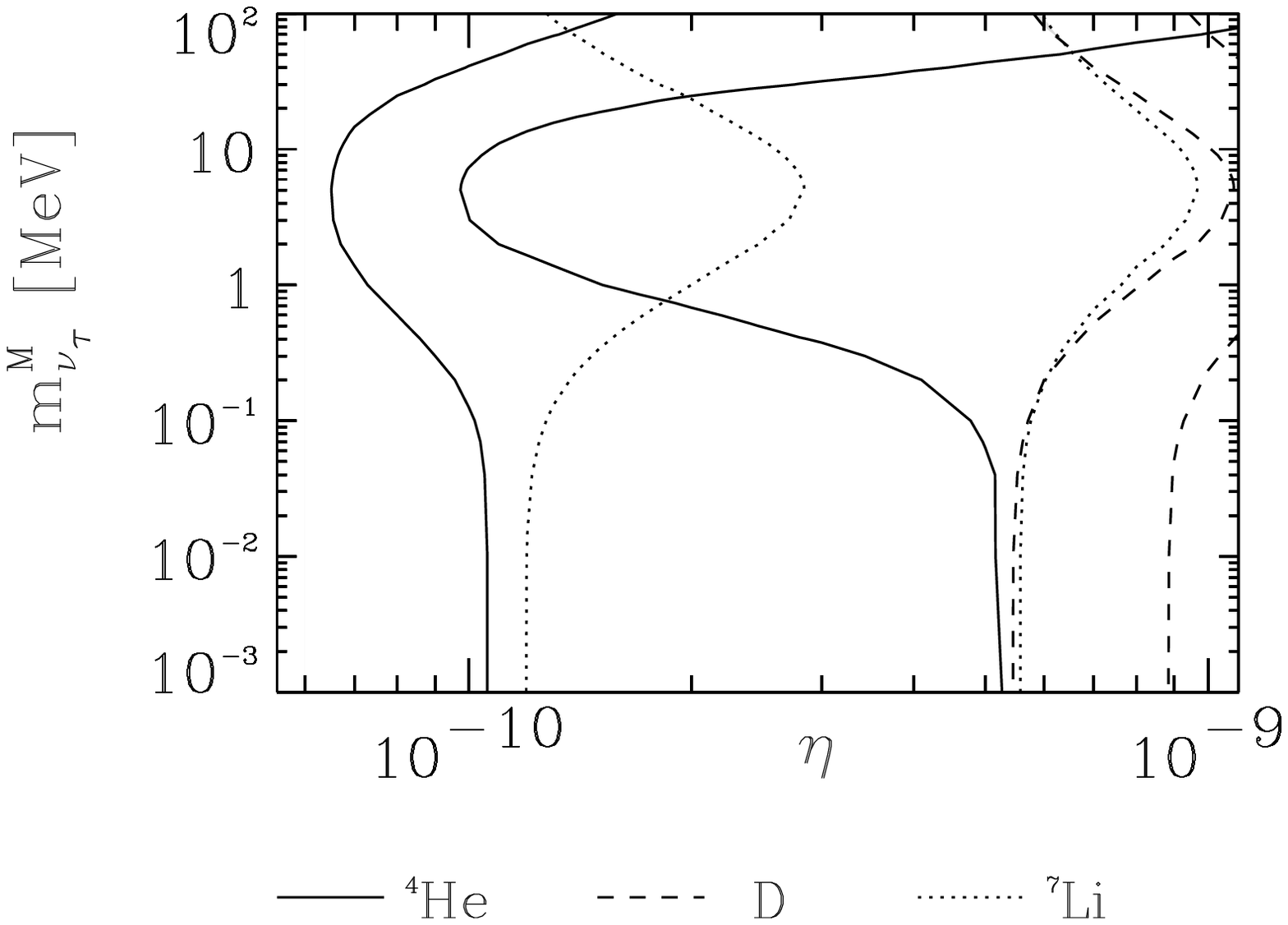,width=8.8cm}
\psfig{figure=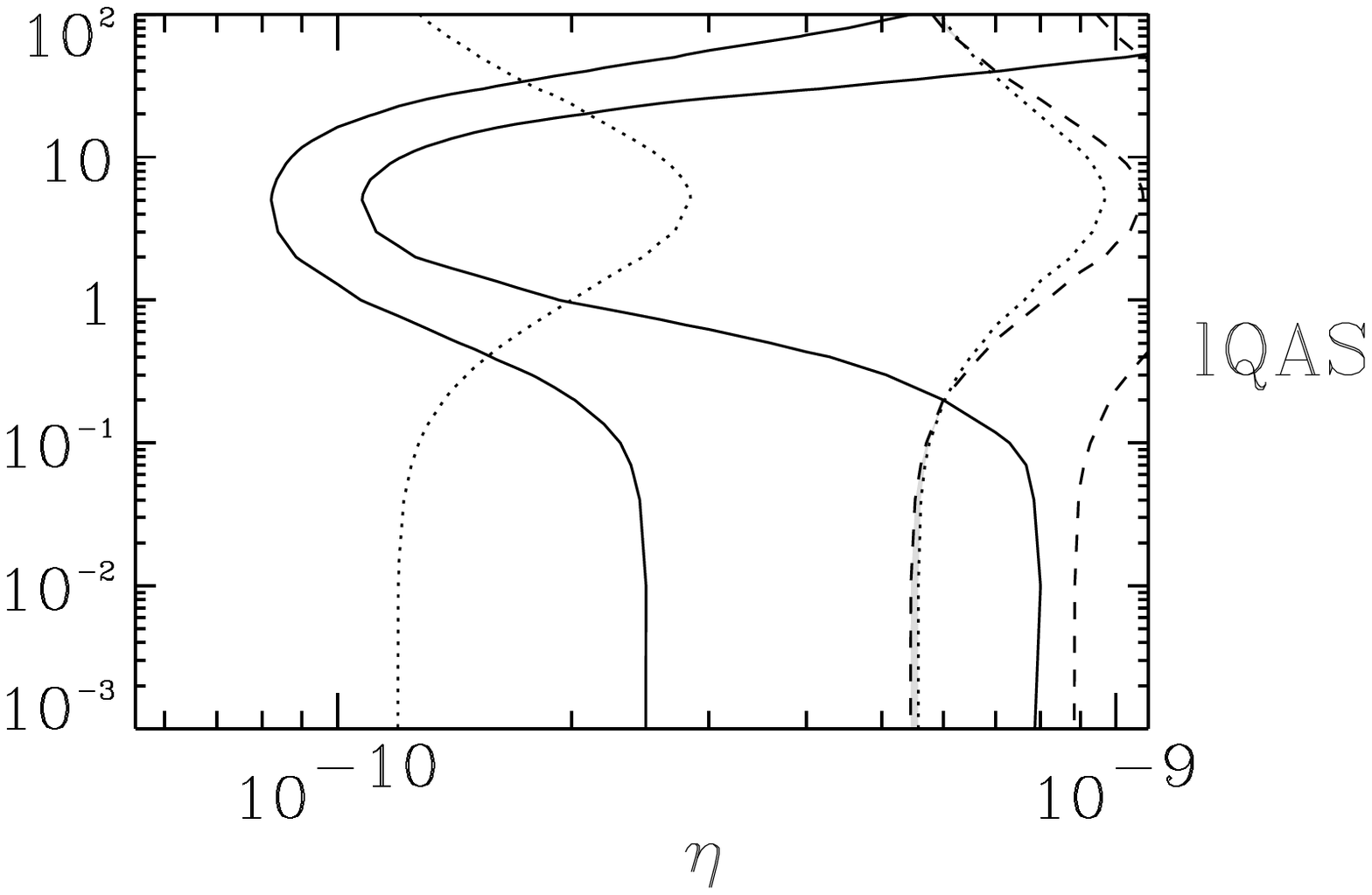,width=8.8cm}}
%\picplace{25cm}
\caption{Isoabundance contours for the light elements if the neutrino
mass is taken to be of Majorana type. In each panel, the contours have
been drawn for a specific combination of observational data which are
denoted according to the labels given in Table~\protect\ref{t:SBBN}.
(Left column: low $\he4$. Right column: high $\he4$.  Top row:
deuterium from ISM. Middle row: high deuterium from QAS. Bottom row:
low deuterium from QAS.)  The concordance regions for $\eta$ and
$m_{\nu_\tau}$ are shaded.
\label{exclmaj}}
\end{figure*}

\begin{figure*}
\centerline{\psfig{figure=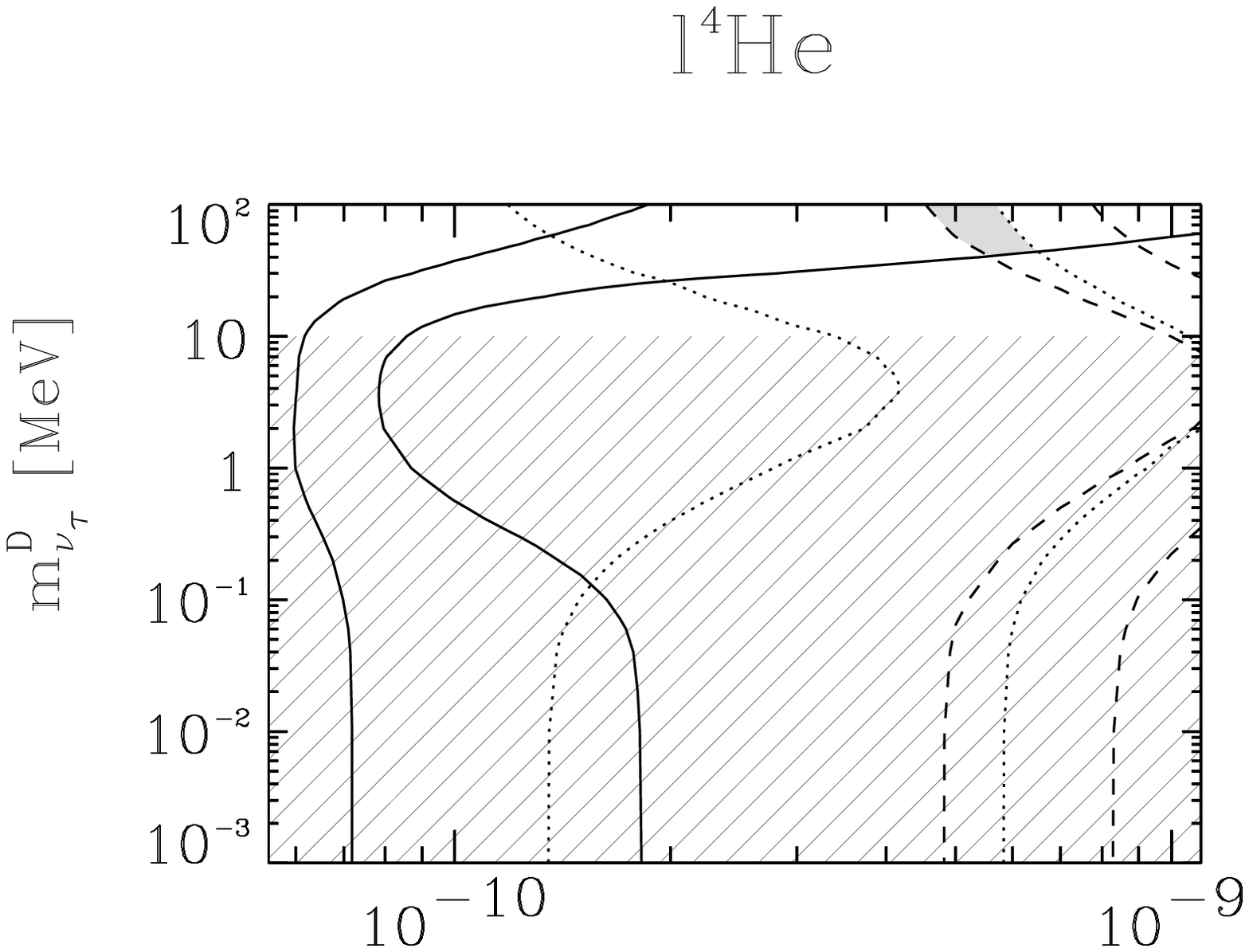,width=8.8cm}
\psfig{figure=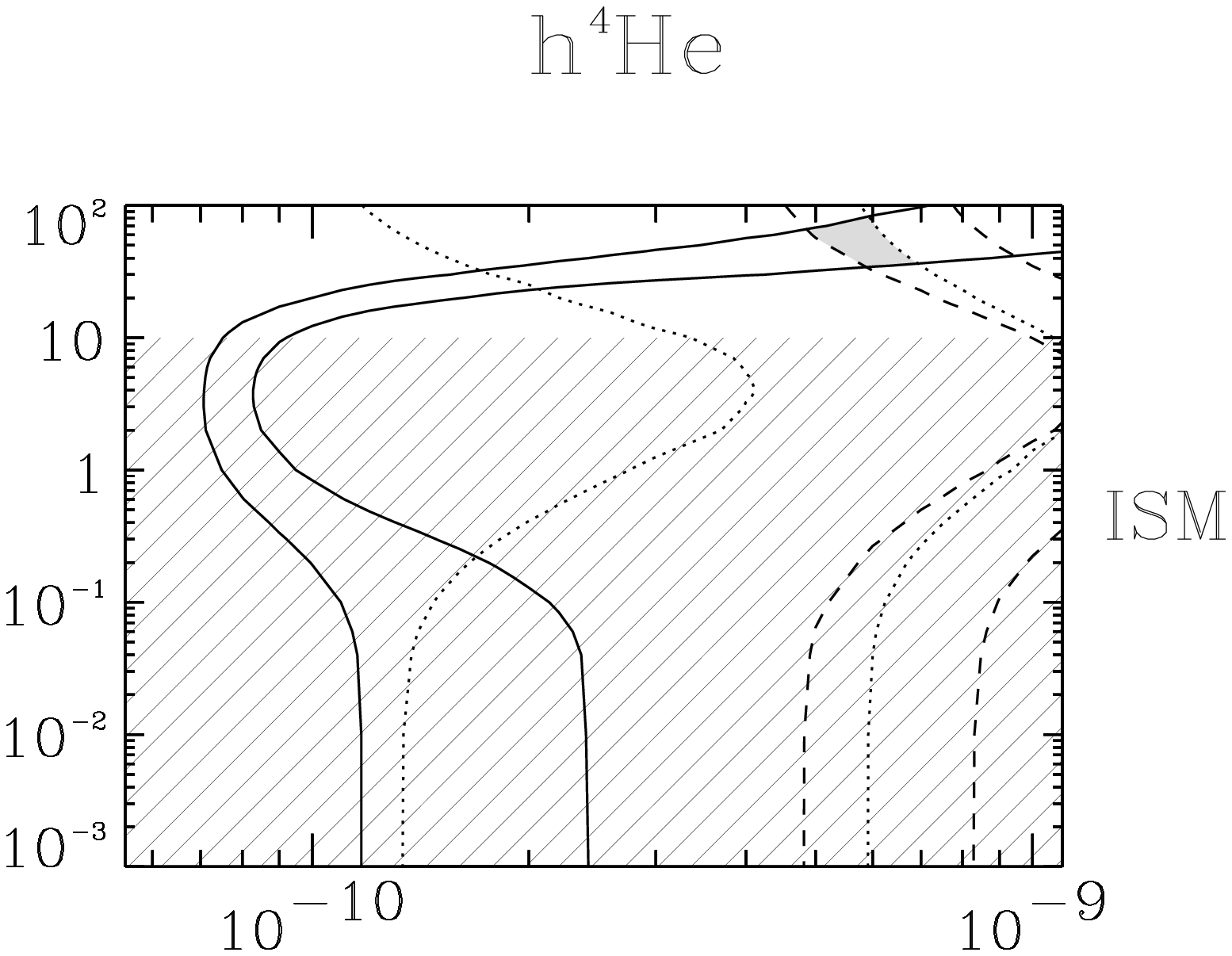,width=8.8cm}} 
\vskip-1cm
\centerline{\psfig{figure=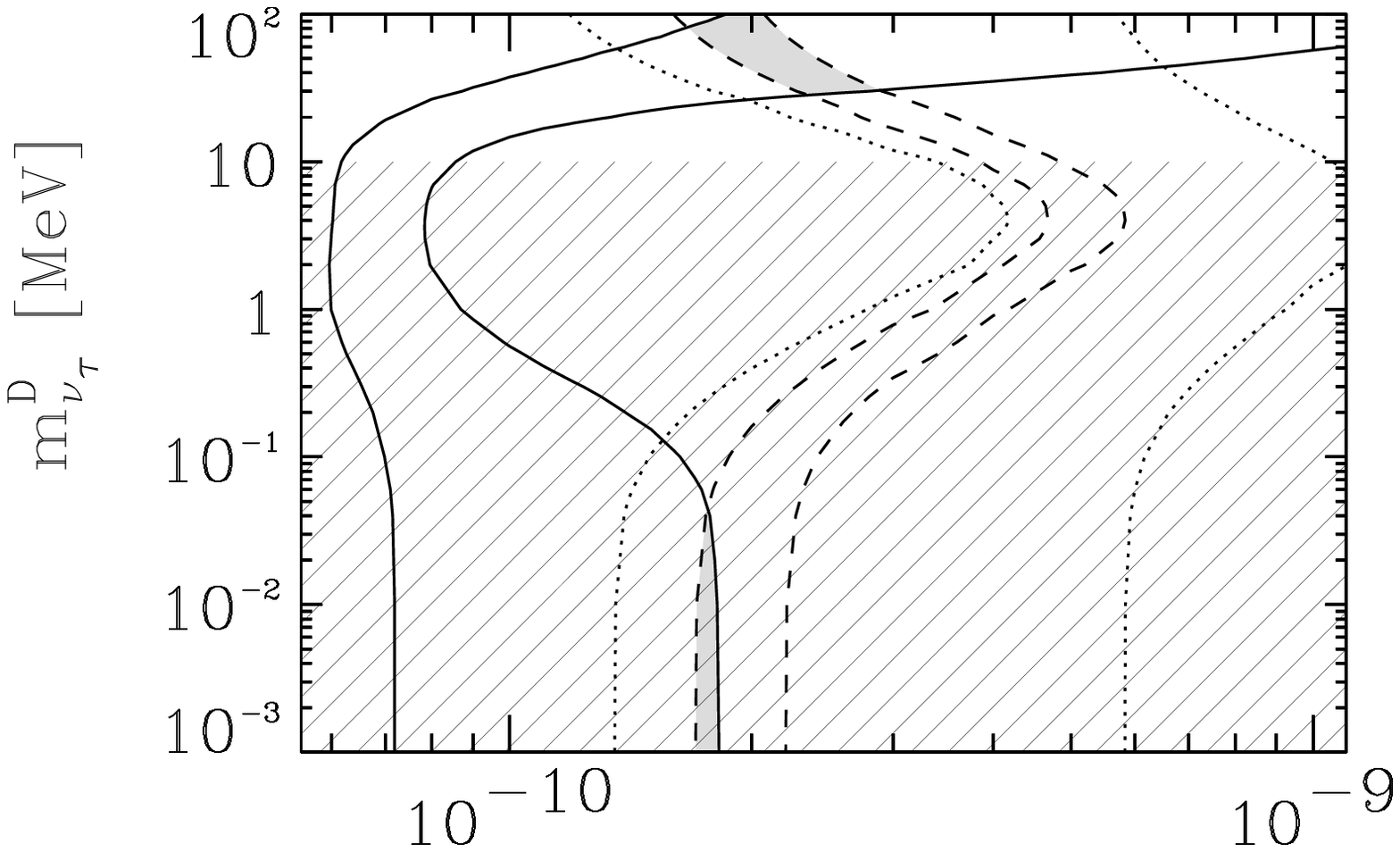,width=8.8cm}
\psfig{figure=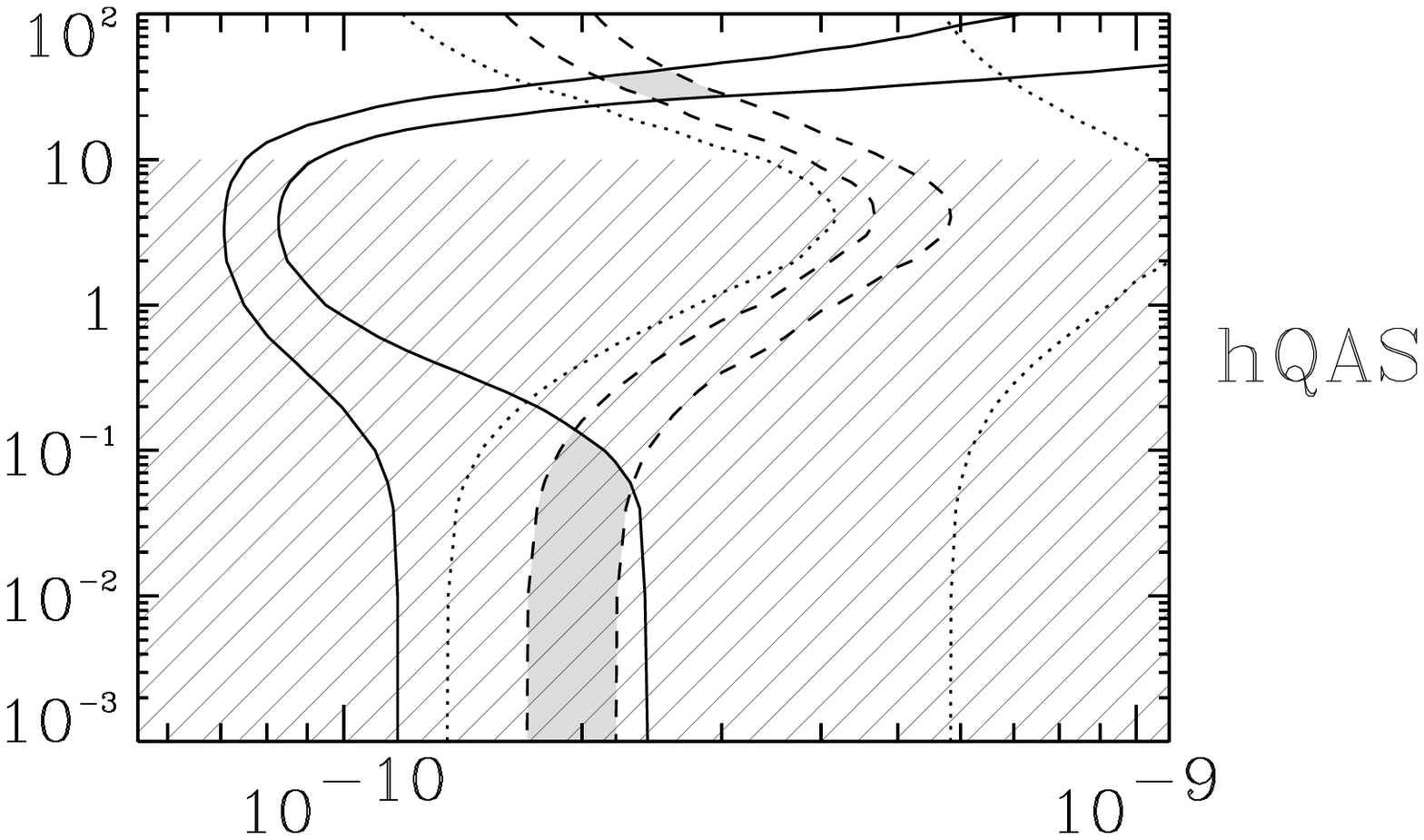,width=8.8cm}} 
\vskip-1cm
\centerline{\psfig{figure=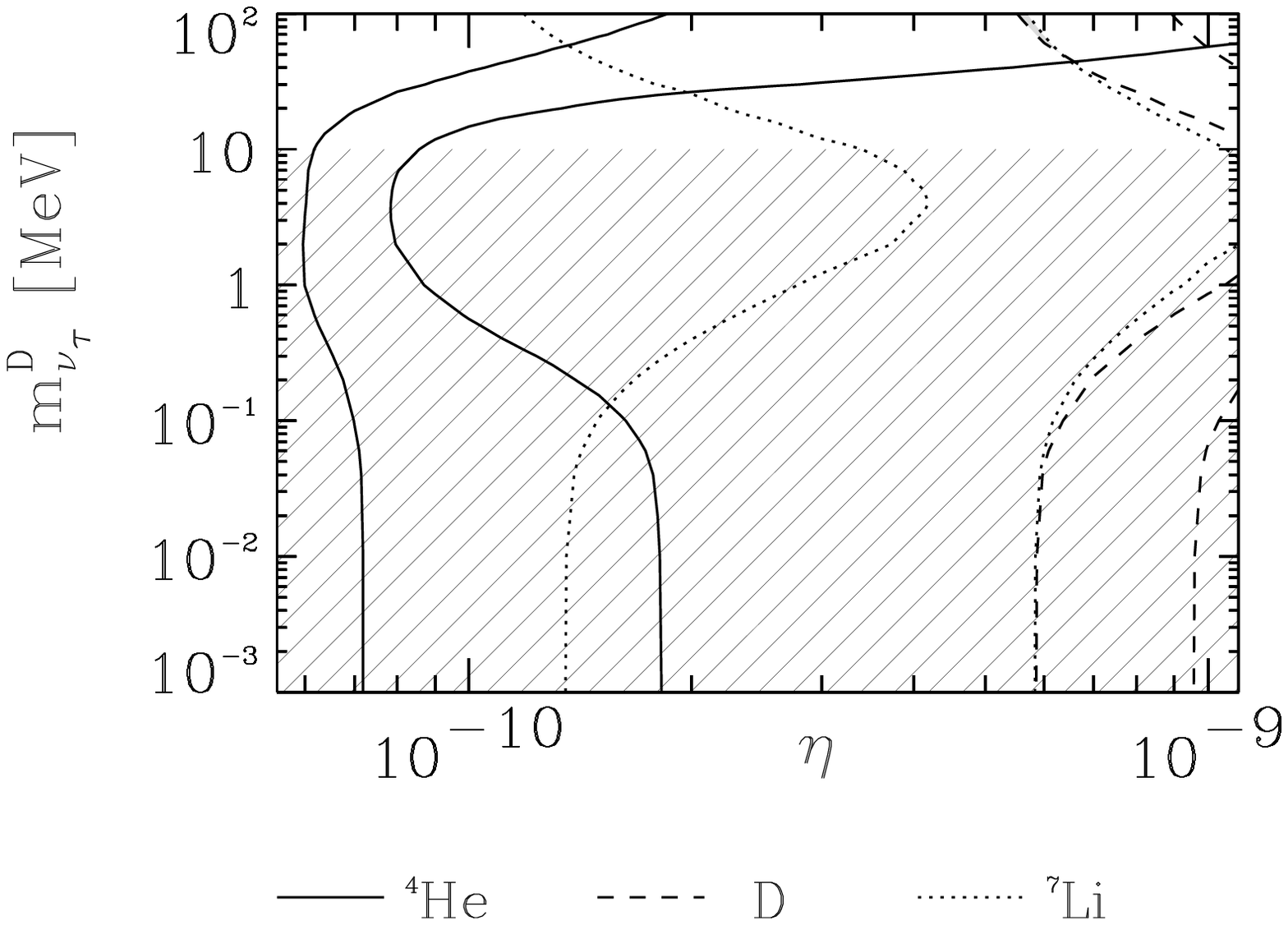,width=8.8cm}
\psfig{figure=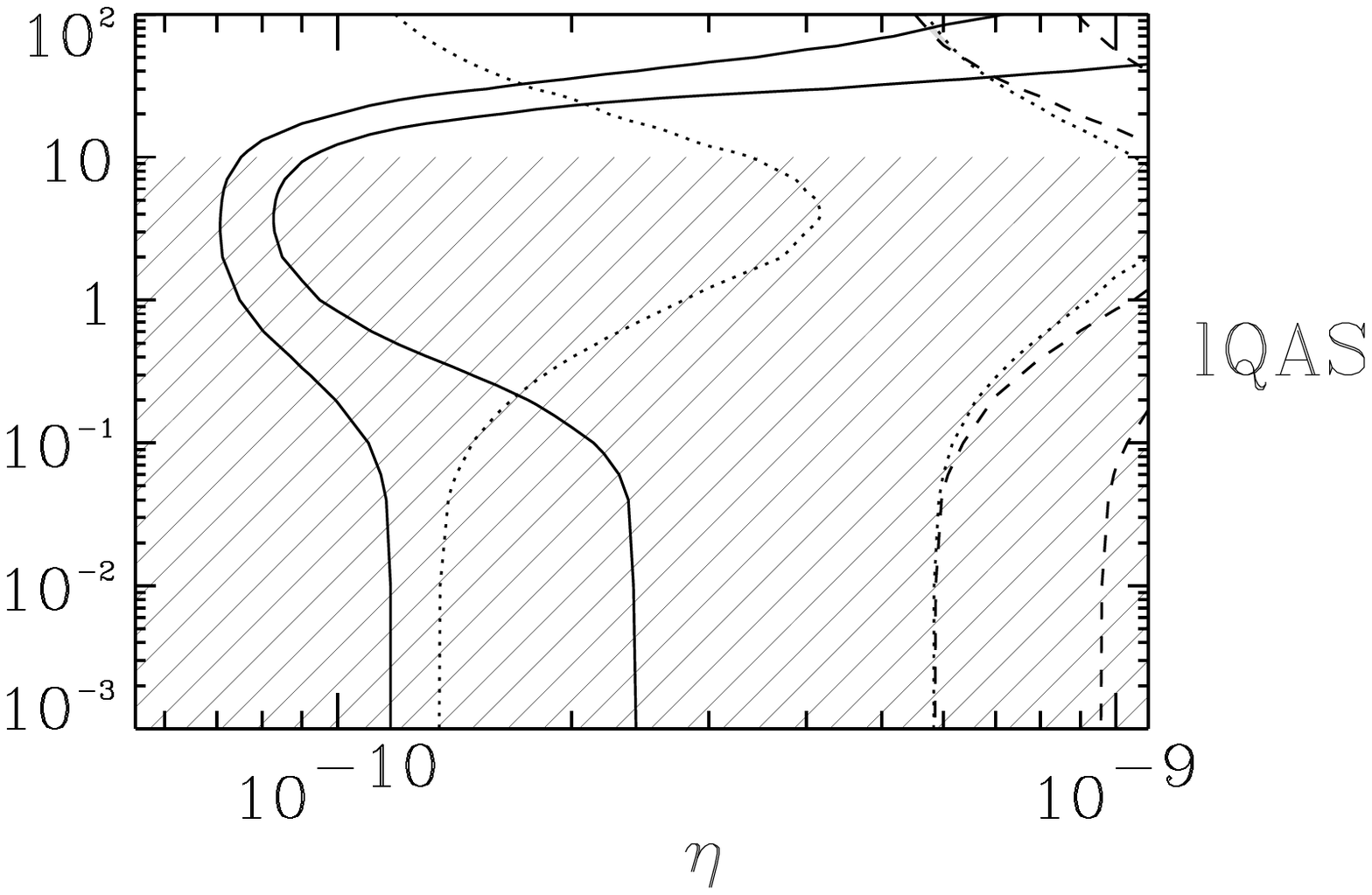,width=8.8cm}} 
%\picplace{9cm}
\caption{Isoabundance contours as in Fig.~\protect\ref{exclmaj} if the
neutrino mass is of Dirac nature. In the hatched area the results are
not quantitatively correct because of our ``helicity approximation''.
\label{excldir}}
\end{figure*}

\subsection{Validity of the approximated Boltzmann equation}

The $N_{\rm eff}$ parameter remains very useful to compare our results
with those of other authors who have dropped some or all of the
simplifying assumptions made in Sect.~\ref{s:massnu} to calculate the
neutrino freeze-out.  \citeasnoun{FKO:97} used the integrated version
of the Boltzmann equation, but with correct Fermi-Dirac statistics for
the neutrinos. Furthermore, they did not assume thermodynamic
equilibrium for the electron neutrinos, but solved the coupled
Boltzmann equation for all three neutrino families and took into
account the influence of the increased $\nu_{\rm e}$ number density on
the weak reaction rates. This changes the maximum value of $N_{\rm
eff}$ by about 0.5, but the limits on the allowed \tnu\ mass are
hardly affected, because below $N_{\rm eff}=4$ the results practically
coincide.

\indent\citename{HM:96} (1996a,b) \nocite{HM:96b} solved the full
Boltzmann equation for all neutrino species without any simplifying
assumptions. In addition they calculated the influence of the
deviation from equilibrium of the electron neutrino distribution
function on the weak reaction rates.  Their results deviate at most by
0.2 neutrino families compared to \citeasnoun{FKO:97}.

\indent \citeasnoun{DPV:96} solved the integrated Boltzmann equation
with the method of the pseudo chemical potential only for the \tnus,
but calculated the deviation from equilibrium of the electron neutrino
distribution function on the weak reaction rates. In this paper they
do not present values for $N_{\rm eff}$, but only for $\Delta N_{\rm
eff}$ relative to their earlier paper (\cite{DR:93}) for the results
without $\nu_{\rm e}$-heating. There, very large values for $N_{\rm
eff}$ were shown with $N_{\rm eff, max}^{\rm Dirac}\approx 11$ and
$N_{\rm eff, max}^{\rm Majorana}\approx 8$.  Their results for $\Delta
N_{\rm eff}$ are opposite to those of the other groups. They argue
that this is because \citeasnoun{FKO:97} did not take into account the
deviation from equilibrium for the distribution function of the
electron neutrinos, but \citeasnoun{Ka:96} showed in an update to
\citeasnoun{FKO:97} that this would change their bounds by less than
half a neutrino family.

In summary, the modifications of $N_{\rm eff}$ brought about by a more
complete kinetic treatment of neutrino freeze-out and the weak
interaction rates appear to be small enough to justify our simple
treatment with the integrated Boltzmann equation.

In the study of \citeasnoun{KKKSSW:94} the $N_{\rm eff}$ parameter was
not introduced, and they performed a rather complete numerical
treatment of the $\nu_\tau$ kinetics. We have constructed a contour
plot with the observational data used by \citeasnoun{KKKSSW:94} in
order to be able to compare directly with their Figure~5(e) which
refers to the stable Majorana neutrino case. Our contours agree with
theirs better than we would have expected, corroborating that our
simple approximation is entirely justified for the task at hand. 

\subsection{Concordance Regions}

In Figs.~\ref{exclmaj} and \ref{excldir} we present our results as
isoabundance contours in the plane of baryon density vs.\ neutrino
mass. In each panel, contours are drawn for $^4$He (solid), deuterium
(dashed) and $^7$Li (dotted).  The contour levels correspond to the
adopted observational ranges listed in Table~1. Each panel corresponds
to a specific combination of observational results as described in the
figure. Concordance regions in the $\eta$-$m_{\nu_\tau}$-plane 
are shaded.  In the Dirac case (Fig.~\ref{excldir}) the area
below about $10\,\rm MeV$ has been hatched because there our results are
not quantitatively meaningful because of our helicity approximation.
Qualitatively, the Dirac and Majorana cases, of course, yield rather
similar results.

The Dirac \tnu\ results have to converge 
to the ones for the Majorana 
case in the small mass regime ($m_{\nu_{\tau}} \lsim$ 1 MeV),
since the ``wrong-helicity'' states 
are not populated there. In the intermediate mass range from 1 -- 10 
MeV our approximation still overestimates the contribution of 
the Dirac \tnu\ energy density to the cosmic expansion rate; the 
correct results for the light element abundances lie somewhere between 
the two cases .(For the $\he4$ abundance in the presence of a massive 
Dirac \tnu\ with the correct treatment of the helicity states see  
\cite{FKO:97}.) The correct treatment of both helicity
states does not open  up new concordance regions in this mass range.

The most recent experimental limit on the $\tau$ neutrino mass is
$m_{\nu_\tau} < 18 $ MeV (ALEPH--collaboration, \cite{Pa:96}), somewhat
lower than the previously published limit (ALEPH--collaboration,
\cite{BCD:95}). 
Therefore, the concordance regions above this mass limit are already
ruled out. 

If the classic ISM determination of the primordial deuterium abundance
were correct after all (top row in Figs.~\ref{exclmaj} and
\ref{excldir}) we would be left with a concordance region in the
low-mass regime; in fact the neutrino mass may well be zero in this
case.  Even though the concordance region does not show up in the
Dirac case because of our helicity approximation, it would be there
because for low-mass neutrinos the Dirac and Majorana cases are
equivalent.
The concordance region 
is almost independent of which $^4$He result is used because of the
constraints provided by the deuterium and lithium contours. 
A mass range between a few
hundred keV and few tens of MeV is excluded. 

If the high QAS deuterium abundance is correct (middle row
in Figs.~\ref{exclmaj} and \ref{excldir}), the result depends on the
assumed $^4$He abundance. If the low value is right, the situation is
similar to that discussed in the previous paragraph, except that the
inferred cosmic baryon abundance is smaller by a factor of 3. If the
high helium 
abundance is the true value, then in the Majorana case there remains 
a small concordance region which actually requires a neutrino mass
somewhat below $1\,\rm MeV$. In the Dirac case, the situation is
unclear because of our helicity approximation. 

If the low QAS deuterium abundance is taken to be the primordial one 
 (bottom row in
Figs.~\ref{exclmaj} and \ref{excldir}), there is essentially no
concordance region whatsoever, independently of the neutrino
mass. This situation would constitute a true ``crisis for BBN'' as
there would be no consistency between observations and calculations.
Some other novel ingredient would be needed besides a massive
neutrino.

\section{Summary}
\label{summ}

We have studied the influence of a massive \tnu\ in the MeV range on the
primordial abundances of the light elements. Several recent studies
indicate that our simple approximate treatment of the Boltzmann
collision equation for the neutrinos is quite adequate. We have
focussed on a comparison between the calculated light-element
abundances with current observational data by virtue of contour plots
in the $\eta$-$m_{\nu_\tau}$-plane. This method avoids the use of an
effective number of neutrino families $N_{\rm eff}$ which cannot be
defined simultaneously for all light element abundances.

We find that the existence and location of concordance regions where
all observations can be reproduced depends dramatically on the
assumption which of the abundances currently offered as the ``observed
primordial'' ones are actually correct. Depending on
ones choice or judgment one may find that BBN is in crisis (no
concordance region at all), that a nonvanishing 
neutrino mass is required, or that there are concordance regions even
for a vanishing neutrino mass. 

In each case one finds that a neutrino
mass between the experimental limit of 18~MeV and about 1~MeV (and
a bit below) is ruled out. In this regard our conclusions agree with
those of previous authors. In spite of the great uncertainty of the
observational situation, this conclusion is rather stable
because of the opposite curvature of the deuterium and the helium
contours which peak in opposite $\eta$-directions for neutrino masses of a
few MeV. 

Still, given the unclear observational situation one cannot be
confident in the reliability of such limits. After all, if there were
no concordance region whatsoever so that a new ingredient for BBN
would be required, the possible impact of this new physics on the
present results is entirely unknown. One clearly has to wait until the
mutual inconsistencies of the observationally inferred abundances 
have been sorted out before one can arrive at far-reaching cosmological
conclusions
about the properties of elementary particles. 

\vspace*{.5cm}
\begin{acknowledgements}
This work was supported by the ``Sonderforschungsbereich 375-95 f\"ur
Astro-Teilchenphysik'' der Deutschen Forschungsgemeinschaft. J.R.\ has
submitted this work to the Ludwig-Maximilians-Universit\"at M\"unchen
in partial fulfillment of the requirements for his {\it Diplom\/}
(Master of Science) degree. 
We thank A.~Dolgov, G.~Steigman, and K.~Kainulainen for critical comments
on earlier versions of this manuscript. 
\end{acknowledgements}

%%%%%%%%%%%%%%%%%%%%%%%%%%%%%%%%%%%%%%%%%%%%%%%%%%%%%%%%%%%%%%%%%%%%%%%%%%%%%%%
%% Appendix %%%%%%%%%%%%%%%%%%%%%%%%%%%%%%%%%%%%%%%%%%%%%%%%%%%%%%%%%%%%%%%%%%%
%%%%%%%%%%%%%%%%%%%%%%%%%%%%%%%%%%%%%%%%%%%%%%%%%%%%%%%%%%%%%%%%%%%%%%%%%%%%%%%

\appendix
\section{Annihilation rates for massive \tnus}\label{s:rates}

The reaction rate for \tnus\ annihilating to fermions is given by the
thermal average over the cross section $\sigma$ times the relativistic
invariant relative velocity $v$ (\cite{GG:91})
\begin{eqnarray}\label{a:rates}
\left\langle\sigma v %_{\mbox{\scriptsize M\o ller}}
\right\rangle&=&\frac{1}{4m_{\nu_\tau}^4T
K_2^2(m_{\nu_\tau}/T)}\\\nonumber
&&\int_{4m^{2}_\tau}^\infty  \tot s\;
\sigma v %_{\mbox{\scriptsize M\o ller}}
E_{\nu_\tau}E_{\bar{\nu}_\tau}
\sqrt{s-4m_{\nu_\tau}^2}
 K_1(\sqrt{s}/T).
\end{eqnarray}
 $K_i$ is the modified Bessel function of the second kind of order $i$
as defined in \citeasnoun{AS:72}, $E_i$ are the particle energies,
$s=(P_{\nu_\tau}+P_{\bar{\nu}_\tau})^2$ is the center of mass energy,
and $P$ are the four momenta.  We borrowed the expression $\sigma v
E_{\nu_\tau}E_{\bar{\nu}_\tau}$ from \citeasnoun{DR:93}. The
expressions given there are misprinted in several ways as confirmed by
the authors (private communication). 
%\begin{minipage}{\columnwidth}
The correct expressions must read for
Dirac neutrinos
\begin{eqnarray}\label{a:cross:dir}
\parbox{10pt}{$\sigma^{\rm D} v  %_{\mbox{\scriptsize M\o ller}}
E_{\nu_\tau}E_{\bar{\nu}_\tau}$}&&\\
=&&\displaystyle{\frac{1}{4}\frac{1}{g_{{\nu}_\tau}g_{\nu_\tau}}\sum_a\sum_{\rm
Spins}\int|M_{\nu_\tau\bar{\nu}_\tau\leftrightarrow a\bar{a}}|_{\rm
D}^2}\nonumber\\\nonumber
&&\displaystyle{(2\pi)^4
\delta^4(P_{\nu_\tau}+P_{\bar{\nu_\tau}}- P_{a_1} - P_{a_2})
\frac{\tot^3{\bf p}_a}{2E_a(2\pi)^3}
\frac{\tot^3{\bf p}_{\bar{a}}}{2E_{\bar{a}}(2\pi)^3}}\\\nonumber
=&&\displaystyle{\frac{G_{\rm
F}^2 w}{4\pi}\sum_a\left[\frac{1}{6}
\left(C_{{\rm V},a}^2+C_{{\rm A},a}^2\right)\left(s-
m_a^2-m_{\nu_\tau}^2\frac{w^2}{s^2}\right)\right.}\\\nonumber
&&\displaystyle{\left.+\;\frac{m_a^2}{s}\left(C_{{\rm V},a}^2-C_{{\rm
A},a}^2\right)
\left(\frac{s}{2}-m^2_{\nu_\tau} \right)\right]},\nonumber
\end{eqnarray}
%\end{minipage}
and for Majorana neutrinos
\begin{eqnarray}\label{cross:maj}
\parbox{10pt}{$\sigma^{\rm M} v
E_{\nu_\tau}E_{ {\nu}_\tau}$}&&\\
=&&\displaystyle{\frac{1}{4}\frac{1}{{g_{ {\nu}_\tau}g_{\nu_\tau}}}\sum_a
\sum_{\rm Spins}\int\frac{1}{2}|M_{\nu_\tau {\nu}_\tau\leftrightarrow
a\bar{a}}|_{\rm M}^2}\nonumber\\
\times&&\displaystyle{(2\pi)^4
\delta^4(P_{\nu_\tau}+P_{\bar{\nu_\tau}}- P_{a_1} - P_{a_2})
\frac{\tot^3{\bf p}_a}{2E_a(2\pi)^3}
\frac{\tot^3{\bf p}_{\bar{a}}}{2E_{\bar{a}}(2\pi)^3}}\nonumber\\
=&&\displaystyle{\frac{G_{\rm F}^2w}{2\pi s} \sum_a
\left\{\left(C_{{\rm V},a}^2+C_{{\rm A},a}^2\right)m_{\nu_\tau}^2
\left(m_a^2-\frac{s}{2}\right)\right.}\nonumber\\
+&&\displaystyle{\left.\left(C_{{\rm A},a}^2-C_{{\rm V},a}^2\right)m_a^2
\left(3m_{\nu_\tau}^2-\frac{s}{2}\right)\right.}\nonumber\\
+&&\displaystyle{\frac{1}{3s}\left(C_{{\rm V},a}^2+C_{{\rm
A},a}^2\right)} 
\displaystyle{\left[\left.\frac{w^2}{2}\left(\frac{s}{2}-m_{\nu_\tau}^2
\right)+\frac{s^2}{4}
\left(s+2m_a^2\right)\right]\right\}}.\nonumber
\end{eqnarray}
Here, {\bf p}$_i$ are the three momenta, $G_{\rm F}$ the
Fermi-constant, $m_a$, $C_{{\rm V},a}$, and $C_{{\rm A},a}$ are the
mass and the vector and axial couplings for the final state
fermions. Further we have defined $w=(s^2-4sm_a^2)^{1/2}$.  The
squared matrix
element $|M_{\nu_\tau\bar{\nu}_\tau\leftrightarrow a\bar{a}}|^2$ is
summed over the kinematically allowed reaction channels
($\nu_{\tau}\bar{\nu}_{\tau}\to e^+e^-$, $\nu_{e}\bar{\nu}_{e}$,
$\nu_{\mu}\bar{\nu}_{\mu}$) and averaged over the spins.

%%%%%%%%%%%%%%%%%%%%%%%%%%%%%%%%%%%%%%%%%%%%%%%%%%%%%%%%%%%%%%%%%%%%%%%%%%%%%%%
%% Bibliography %%%%%%%%%%%%%%%%%%%%%%%%%%%%%%%%%%%%%%%%%%%%%%%%%%%%%%%%%%%%%%%
%%%%%%%%%%%%%%%%%%%%%%%%%%%%%%%%%%%%%%%%%%%%%%%%%%%%%%%%%%%%%%%%%%%%%%%%%%%%%%%


\begin{thebibliography}{50}

\bibitem[\protect\astroncite{Abramowitz \& Stegun}{1972}]{AS:72}
Abramowitz M., Stegun I.S. (eds.), 1972, {\em Handbook of mathematical
  functions\/}, Dover Publications Inc., New York

\bibitem[\protect\astroncite{Buskulic et~al.}{1995}]{BCD:95}
Buskulic D. et~al., 1995, {\em Phys. Lett. B\/} 349, 585, (ALEPH Collaboration)

\bibitem[\protect\astroncite{Cardall \& Fuller}{1996}]{CF:96}
Cardall C., Fuller G.M., 1996, {\em ApJ\/} 472, 435

\bibitem[\protect\astroncite{Carswell et~al.}{1994}]{CRWCW:94}
Carswell R.F., Weymann R.J., Cooke A.J., Webb K.J., 1994, {\em Mon. Not. R.
  Astron. Soc.\/} 268, L1

\bibitem[\protect\astroncite{Charbonnel}{1995}]{Ch:95}
Charbonnel C., 1995, {\em ApJ\/} 453, L51

\bibitem[\protect\astroncite{Copi et~al.}{1995}]{CST:95}
Copi C.J., Schramm D.N., Turner M.S., 1995, {\em Phys. Rev. Lett.\/} 75, 3981

\bibitem[\protect\astroncite{Dearborn et~al.}{1996}]{DST:96}
Dearborn D., Steigman G., Tosi M., 1996, {\em ApJ\/} 465, 887

\bibitem[\protect\astroncite{Deliyannis et~al.}{1990}]{DDK:90}
Deliyannis C.P., Demarque P., Kawaler S.D., 1990, {\em ApJ Suppl. Ser.\/} 73,
  21

\bibitem[\protect\astroncite{Dicus et~al.}{1978}]{DKTW:78}
Dicus D.A., Kolb E.W., Teplitz V.L., Wagoner R.V., 1978, {\em Phys. Rev. D\/}
  17, 1529

\bibitem[\protect\astroncite{Dodelson et~al.}{1994}]{DGT:94}
Dodelson S., Gyuk G., Turner M.S., 1994, {\em Phys. Rev. D\/} 49, 5068

\bibitem[\protect\astroncite{Dolgov \& Kirilova}{1988}]{DK:88}
Dolgov A.D., Kirilova D., 1988, {\em Int. J. Mod. Phys. A\/} 3, 267

\bibitem[\protect\astroncite{Dolgov et~al.}{1996}]{DPV:96}
Dolgov A.D., Pastor S., Valle J., 1996, {\em Phys. Lett. B\/} 383, 193

\bibitem[\protect\astroncite{Dolgov \& Rothstein}{1993}]{DR:93}
Dolgov A.D., Rothstein I.Z., 1993, {\em Phys. Rev. Lett.\/} 71, 476

\bibitem[\protect\astroncite{Edmunds}{1994}]{Ed:94}
Edmunds M.G., 1994, {\em Mon. Not. R. Astron. Soc.\/} 270, L37

\bibitem[\protect\astroncite{Enqvist et~al.}{1997}]{EKMU:96}
Enqvist K., Ker\"anen P., Maalampi J., Uibo H., 1997, {\em Nucl.Phys.B\/} 484,
  403

\bibitem[\protect\astroncite{Fields et~al.}{1997}]{FKO:97}
Fields B.D., Kainulainen K., Olive K.A., 1997, {\em Astropart. Phys.\/} 6, 169

\bibitem[\protect\astroncite{Fields et~al.}{1996}]{FKOT:96}
Fields B.D., Kainulainen K., Olive K.A., Thomas D., 1996, {\em New Astronomy\/}
  1, 77

\bibitem[\protect\astroncite{Galli et~al.}{1995}]{GPFP:95}
Galli D., Palla F., Ferrini F., Penco U., 1995, {\em ApJ\/} 443, 536

\bibitem[\protect\astroncite{Geiss}{1993}]{Ge:93}
Geiss J., 1993, in: {\em Origin and Evolution of the Elements\/} (Edited by
  Prantzos N., Vagioni-Flam E., Cass\'e M.),  p. 89, Cambridge, UK, Cambridge
  University Press

\bibitem[\protect\astroncite{Gondolo \& Gelmini}{1991}]{GG:91}
Gondolo P., Gelmini G., 1991, {\em Nucl. Phys. B\/} 360, 145

\bibitem[\protect\astroncite{Hannestad \& Madsen}{1996{a}}]{HM:96}
Hannestad S., Madsen J., 1996{a}, {\em Phys. Rev. Lett.\/} 76, 2848

\bibitem[\protect\astroncite{Hannestad \& Madsen}{1996{b}}]{HM:96b}
Hannestad S., Madsen J., 1996{b}, {\em Phys. Rev. D\/} 54, 7894

\bibitem[\protect\astroncite{Hata et~al.}{1995}]{HSSTWBL:95}
Hata N. et~al., 1995, {\em Phys. Rev. Lett.\/} 75, 3977

\bibitem[\protect\astroncite{Izotov et~al.}{1997}]{ITL:97}
Izotov Y.I., Thuan T.X., Lipovetsky V.A., 1997, {\em ApJ Suppl. Ser.\/} 108, 1

\bibitem[\protect\astroncite{Kainulainen}{1996}]{Ka:96}
Kainulainen K., 1996, {\em hep-ph/9608215,\/} Talk given at Neutrino 96,
  Helsinki, June 1996

\bibitem[\protect\astroncite{Kawasaki et~al.}{1994}]{KKKSSW:94}
Kawasaki M. et~al., 1994, {\em Nucl. Phys. B\/} 419, 105

\bibitem[\protect\astroncite{Kolb \& Scherrer}{1982}]{KS:82}
Kolb E.W., Scherrer R.J., 1982, {\em Phys. Rev. D\/} 25, 1481

\bibitem[\protect\astroncite{Kolb \& Turner}{1990}]{kt:90}
Kolb E.W., Turner M.S., 1990, {\em The Early Universe\/}, Frontiers in Physics,
  Addison-Wesley, Reading, MA

\bibitem[\protect\astroncite{Kolb et~al.}{1991}]{KTCS:91}
Kolb E.W., Turner M.S., Chakravorty A., Schramm D.N., 1991, {\em Phys. Rev.
  Lett.\/} 67, 533

\bibitem[\protect\astroncite{Linsky et~al.}{1993}]{LBG:93}
Linsky J.L. et~al., 1993, {\em ApJ\/} 402, 694

\bibitem[\protect\astroncite{Linsky \& Wood}{1996}]{LW:96}
Linsky J.L., Wood B.E., 1996, {\em ApJ\/} 463, 254

\bibitem[\protect\astroncite{Malaney \& Mathews}{1993}]{MM:93}
Malaney R.A., Mathews G.J., 1993, {\em Phys. Pep.\/} 229, 145

\bibitem[\protect\astroncite{Miyama \& Sato}{1978}]{MS:78}
Miyama S., Sato K., 1978, {\em Prog. Theor. Phys.\/} 60, 1703

\bibitem[\protect\astroncite{Molaro et~al.}{1995}]{MPB:95}
Molaro P., Primas F., Bonifacio P., 1995, {\em A \& A\/} 295, L47

\bibitem[\protect\astroncite{Olive \& Scully}{1996}]{OS:96}
Olive K.A., Scully S.T., 1996, {\em Int. J. Mod. Phys. A\/} 11, 409

\bibitem[\protect\astroncite{Olive et~al.}{1997}]{OSS:96}
Olive K.A., Skillman E., Steigman G., 1997, {\em ApJ\/} 483, 788

\bibitem[\protect\astroncite{Olive \& Steigman}{1995}]{OS:95}
Olive K.A., Steigman G., 1995, {\em ApJ Suppl. Ser.\/} 97, 49

\bibitem[\protect\astroncite{Passalacqua}{1996}]{Pa:96}
Passalacqua L., 1996, {\em to appear in the proceedings of ``The Fourth
  International Workshop on Tau Lepton Physics (TAU96)'' held at Estes Park,
  Colorado, Sept. 16-19, 1996,\/} available at:\\
  http://tau96.colorado.edu/tau96/ps/passalacqua.ps

\bibitem[\protect\astroncite{Pinsonneault et~al.}{1992}]{PDD:92}
Pinsonneault M.H., Deliyannis C.P., Demarque P., 1992, {\em ApJ Suppl. Ser.\/}
  78, 179

\bibitem[\protect\astroncite{Piskunov et~al.}{1997}]{PWLDA:97}
Piskunov N. et~al., 1997, {\em ApJ\/} 474, 315

\bibitem[\protect\astroncite{Rehm}{1996}]{JR:96}
Rehm J.B., 1996, {\em Der Einflu\ss\ neuer Neutrinoeigenschaften auf die
  Kernsynthese im fr\"uhen Universum\/}, Diplom Thesis,
  Ludwig-Maximilians-Universit\"at, M\"unchen, Germany

\bibitem[\protect\astroncite{Rugers \& Hogan}{1996}]{RH:96}
Rugers M., Hogan C.J., 1996, {\em ApJ\/} 459, L1

\bibitem[\protect\astroncite{Sasselov \& Goldwirth}{1995}]{SG:95}
Sasselov D., Goldwirth D., 1995, {\em ApJ\/} 444, L5

\bibitem[\protect\astroncite{Smith et~al.}{1993}]{SKM:93}
Smith M.S., Kawano L.H., Malaney R.A., 1993, {\em ApJ Suppl. Ser.\/} 85, 219

\bibitem[\protect\astroncite{Songaila et~al.}{1994}]{SCHR:94}
Songaila A., Cowie L.L., Hogan C., Rugers M., 1994, {\em Nature\/} 368, 599

\bibitem[\protect\astroncite{Steigman}{1994}]{St:94}
Steigman G., 1994, {\em Mon. Not. R. Astron. Soc.\/} 269, L53

\bibitem[\protect\astroncite{Steigman et~al.}{1993}]{SFOSW:93}
Steigman G. et~al., 1993, {\em ApJ\/} 415, L35

\bibitem[\protect\astroncite{Tytler et~al.}{1996}]{TFB:96}
Tytler D., Fan X.M., Burles S., 1996, {\em Nature\/} 381, 207

\bibitem[\protect\astroncite{Weiss et~al.}{1996}]{WWD:96}
Weiss A., Wagenhuber J., Denissenkov P.A., 1996, {\em A \& A\/} 313, 581

\bibitem[\protect\astroncite{Yang et~al.}{1984}]{YTSSO:84}
Yang J. et~al., 1984, {\em ApJ\/} 281, 493

\end{thebibliography}
\end{document}